# 70 years of spectroscopy of the photosphere and the solar chromosphere at the Pic du Midi Observatory (1956-2026)

*Jean-Marie Malherbe (jean-marie.malherbe@obspm.fr)*

*Emeritus astronomer at LIRA - Paris Observatory – PSL University*

*18 February 2026 (French version also available at CNRS/HAL)*

**ABSTRACT**

Observations of the solar corona at the Pic du Midi began with Bernard Lyot and his spectro coronagraph installed on the multi-purpose equatorial mount of the Baillaud cupola. It was not until 1956 that domes and instruments specifically dedicated to observations of the photosphere and the solar chromosphere appeared. On the occasion of the International Geophysical Year, a solar spectroscopy laboratory was created to the west of the Pic du Midi, based on two spectrographs of 4 m and 9 m focal length. In 1961 the turret dome appeared to the east of the Pic, later equipped with an 8 m spectrograph. Around 1965, the Baillaud dome finally specialized in the corona with a new table and new spectrographs. At the same time, a revolution in infrastructure took place at the Pic in a few years, which we present as well as the solar spectrographs and their goal.

**INTRODUCTION**

In the 1930s, day-time and night astronomers shared the only dome of the establishment, the Baillaud equatorial. Bernard Lyot (1997-1952) set up there the coronagraph he had invented in Meudon to observe the corona, the Sun's atmosphere $10^6$ times darker than the disk. The bright surface must be masked by an optically optimised instrument combined with pure atmosphere in order to reveal the corona. With the increasing number of projects to be carried out at the Pic, other specialized domes were needed and designed according to the goals of the observing programs. In view of the International Geophysical Year (1957-1958), Raymond Michard (1925-2015), of the Paris Observatory, proposed to Jean Rösch (1915-1999), then director of the Pic since 1947, to set up a spectroscopy laboratory. The aim was to study the spectra of solar flares occurring on the disk with a spectral resolution that would allow the line profiles to be resolved. At the time, the reconstruction of the profiles required the use of large spectrographs (5 to 10 m focal length), which were therefore very dispersive but heavy, due to the size of the sensors (photo plates, 35 or 70 mm films). In 1956, the construction of a new laboratory began to the west of the Pic, which operated regularly for 15 years, more sporadically thereafter. Then, international requirements in terms of spatial resolution, which were increasing, encouraged Jean Rösch to start work on a second equipment, this time better located to the east of the Pic, which was created in 1961 and known as the "turret dome". The exceptional images obtained prompted Zadig Mouradian (1930-2020), from Meudon, who was an expert user of Michard's laboratory, to build a new spectrograph to be attached to the mount of the refractor ; the quality of which finally dethroned the previous installations. This spectrograph is losing activity today because many of astronomers have turned to those of the more recent Canary Island observatories. At the same time, the Baillaud dome was dedicated to spectro coronagraphy, since the stellar astronomers benefited from the 106 cm telescope of the Gentili cupola since 1963, then the 2 m Bernard Lyot telescope (TBL). After describing the context of the remarkable development of the Pic infrastructures in the 50s (section 1), under the auspices of Rösch (see Rösch, 1963), we present the achievements of solar surface spectroscopy in sections 2 and 3, with a brief point on the spectroscopy of the corona in section 4.

**I – THE CONTEXT OF INFRASTRUCTURES IN THE MIDDLE OF XX$^{th}$ CENTURY AT PIC DU MIDI**

Before 1945, the context was that of Figure 1: a single astronomic dome (Baillaud cupola) to the west of the Pic and the original buildings, Nansouty and Vaussenat, then Dauzère in the 1920s (with a tower which was created to establish new laboratories or domes). Access was only on foot. This is how Bernard Lyot worked with the coronagraph from 1930, in the Baillaud dome.

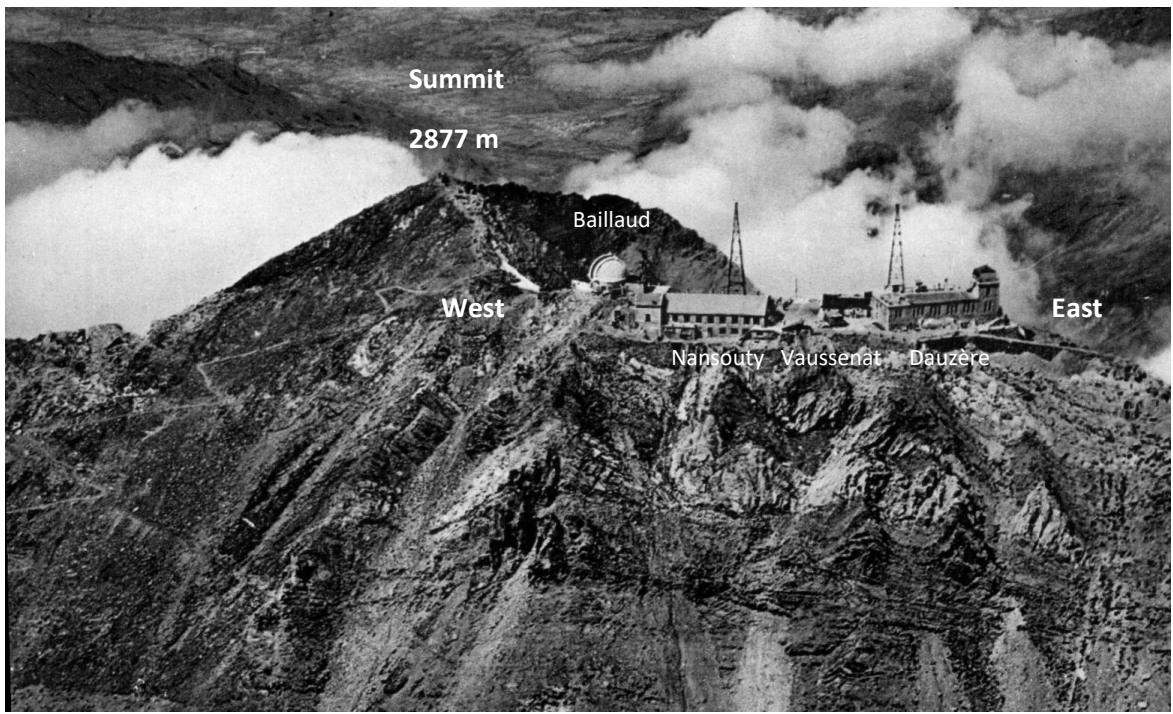

*Figure 1: before 1945, from left to right, Baillaud dome, houses of Toulouse and Nansouty, Vaussenat and Dauzère. The two pylons carry a broadcasting radio antenna (postcard).*

Figure 2 shows the evolution at the beginning of the 50s: the cable car station was completed but it was not operational until 1952. To the east of the Pic, there are two new constructions started in 1947, the "pas de case" (accommodations) adjoining a tower of 10 m in diameter intended to support a future dome. Next to this tower, you can see the arrival of the funicular built in 1949 to bring up to the observatory the 4 tons electromagnet of the Manchester team, which decided to come to the Pic (at Vaussenat's building) in order to study the collisions of cosmic ray particles with the environment, thus used as a nuclear physics tool before the CERN accelerators. Measuring the energy of charged secondary particles requires bending their trajectory by an intense magnetic field created by coils within a Wilson gas expansion chamber. The funicular, on a steep 40° slope, started at the altitude of 2700 m, a point located at the end of the 400 m extension of the road; it was suitable for vehicles (in summer) leading from the Tourmalet pass to the "Auberge des Laquets" at 2650 m. A 50 m railway line then led to the Pic terrace on a slightly inclined plane, clearly visible in front of the Dauzère building (Figure 1) or behind the "pas de case" (Figure 2). The electric line was commissioned at the same time (under 10000 V), it was necessary to power the coils whose field of about 1 T required a high intensity of direct current (therefore a transformer was present). A second team (with Louis Leprince Ringuet from Ecole Polytechnique) came in 1953 with two more powerful Wilson chambers, which were installed in Dauzère's house.

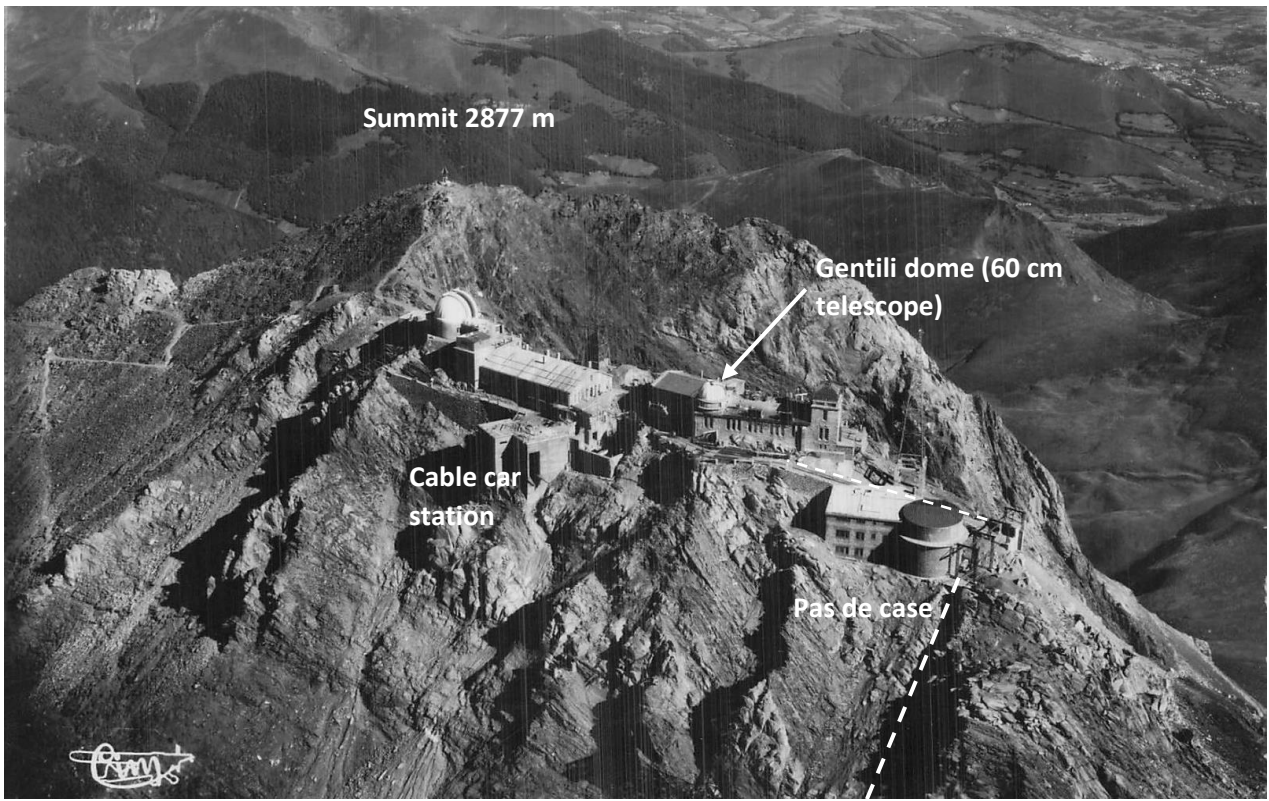

Figure 2: around 1951. The "pas de case" building and the round tower of 10 m in diameter, to the east, were built from 1947 onwards. The portico of the funicular (adjacent to the tower) is from 1949. The cable car station dates from 1950 (cable car in service in 1952). The Gentili dome is from 1947. Postcard.

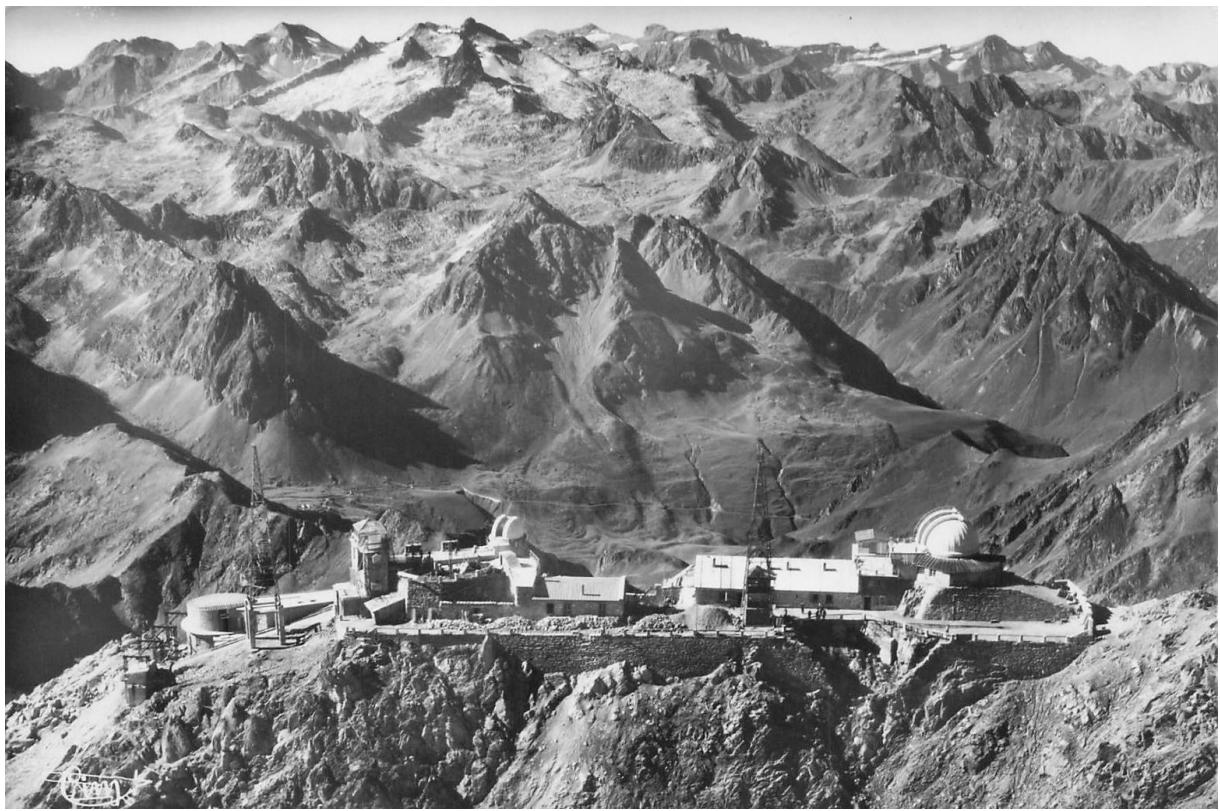

Figure 3: around 1951. View to the south and the pyrenean mountains. Postcard.

Figures 3 to 7 show the Pic du Midi observatory from various points of view in the early 1950s.

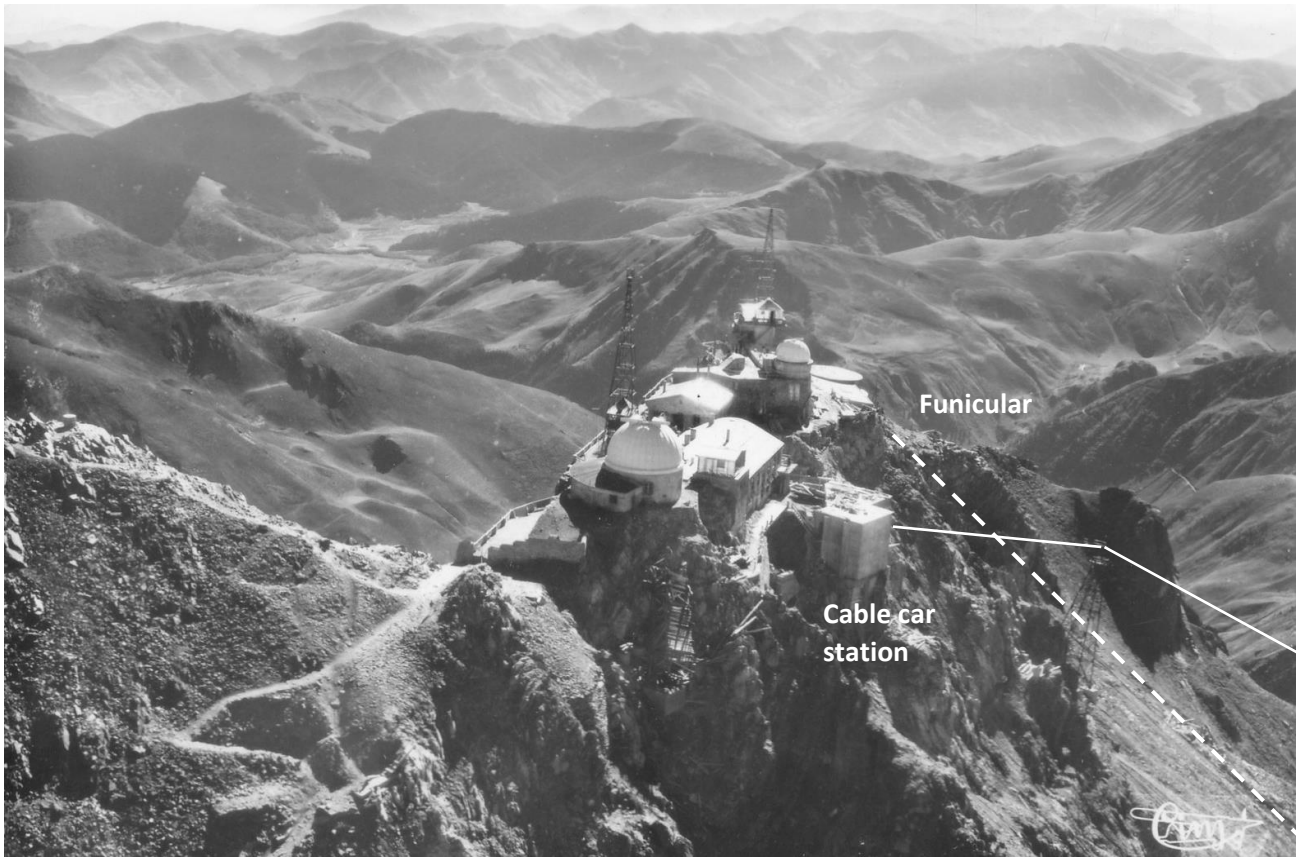

*Figure 4: around 1951. View to the east. Postcard.*

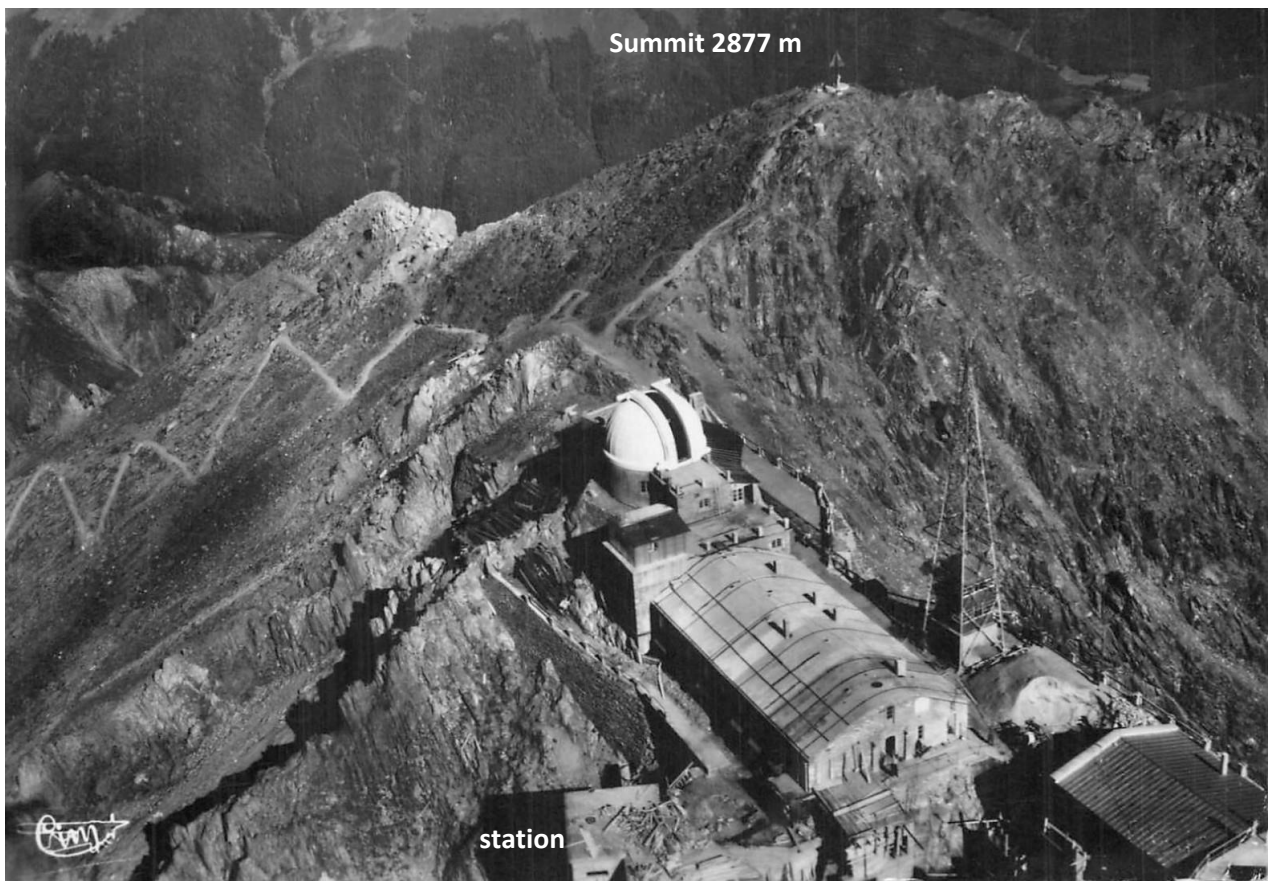

*Figure 5: around 1951. View to the west and the summit. Postcard.*

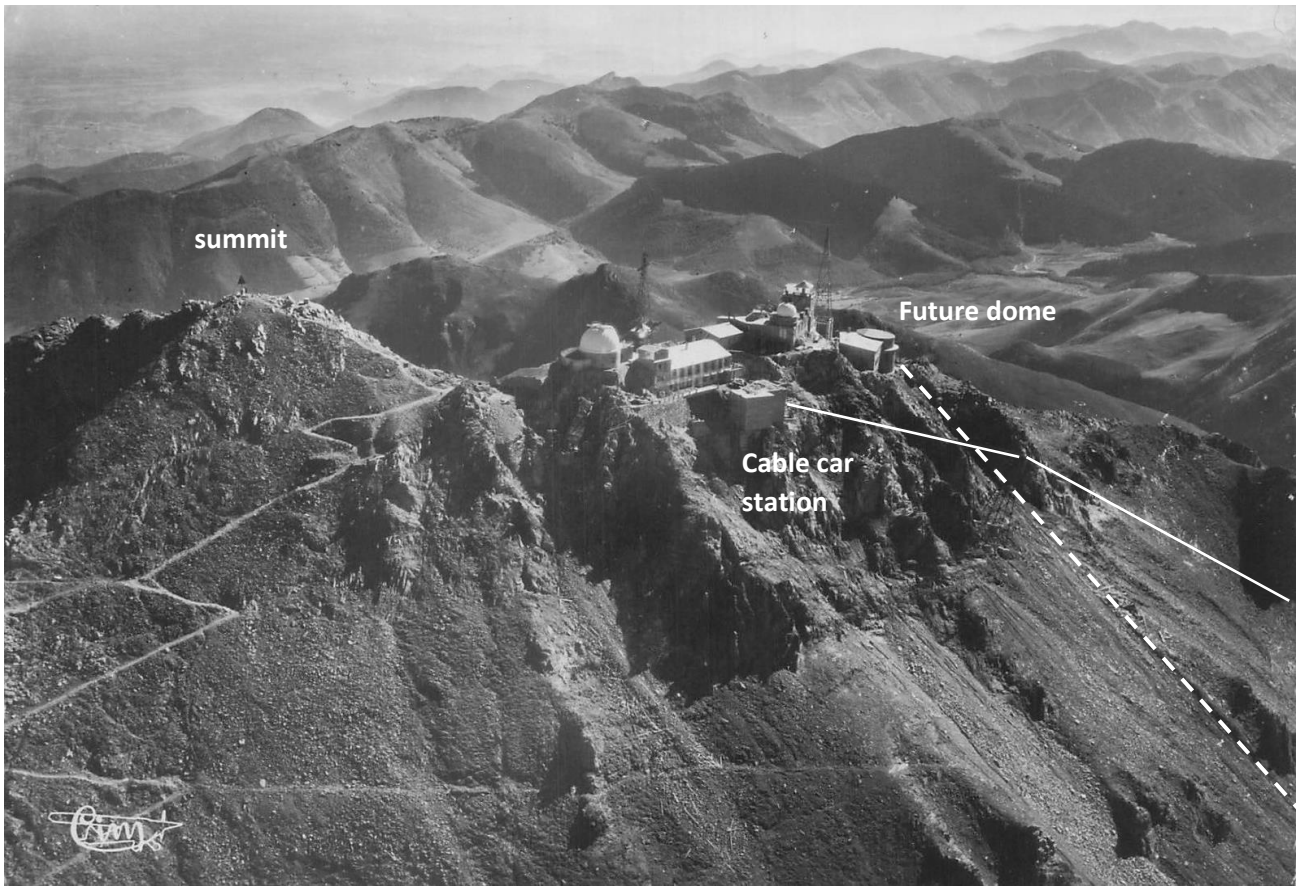

*Figure 6: around 1953. View to the east. We see the funicular of 1949. Postcard.*

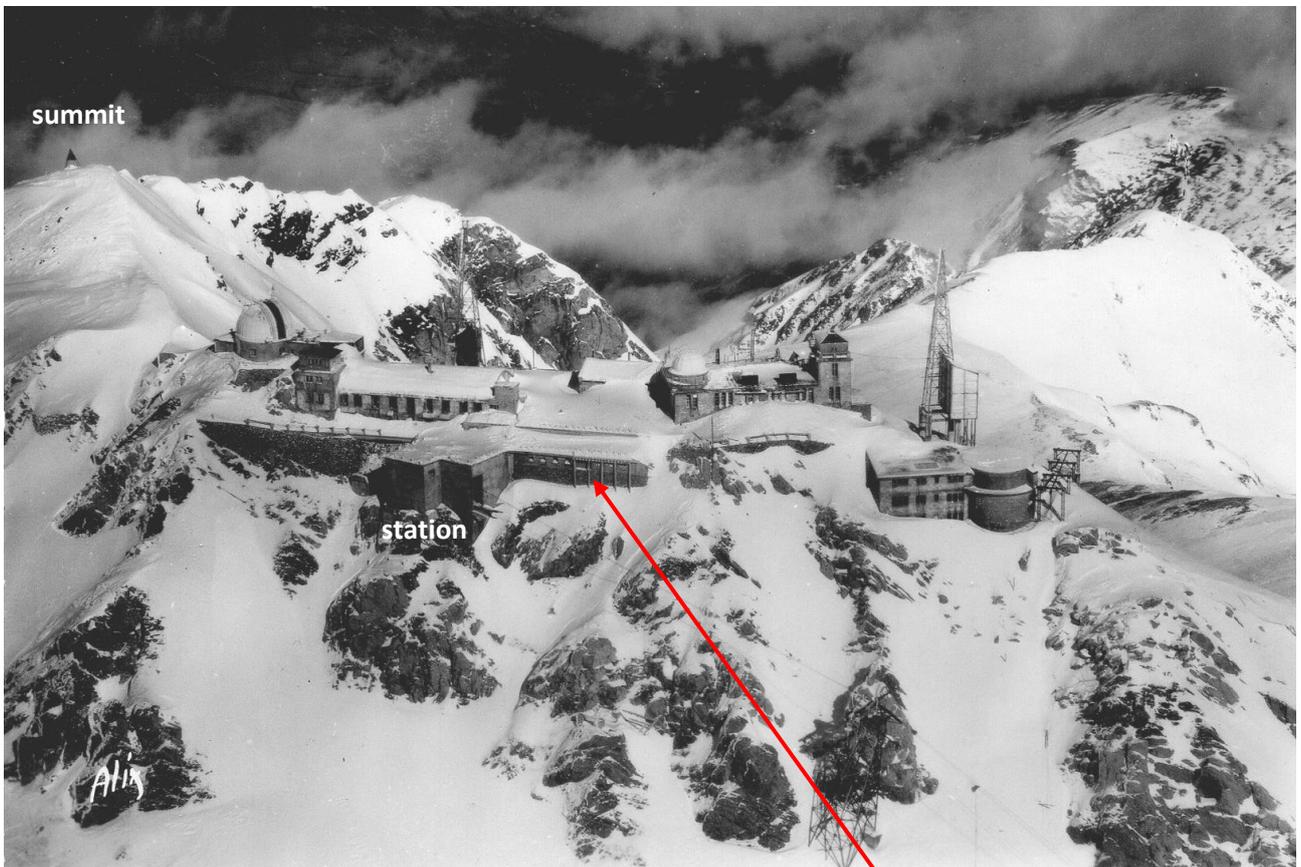

*Figure 7: around 1955. Note the progression of the constructions (the workshop, arrow) between the cable car station and the "pas de case" house to the east. Postcard.*

The construction of the solar spectroscopy laboratory began in the summer of 1956 in the west, just below the Baillaud dome (Figure 8). A coelostat with two mirrors, mounted on a scaffold, supplied it with light (section 2). In 1957, the first television antenna (Figure 8) appeared with 500 W of power and broadcasted the first black and white TV channel with a standard of 819 lines. Preparations were in progress for the installation of the turret dome on the east tower (Figure 8).

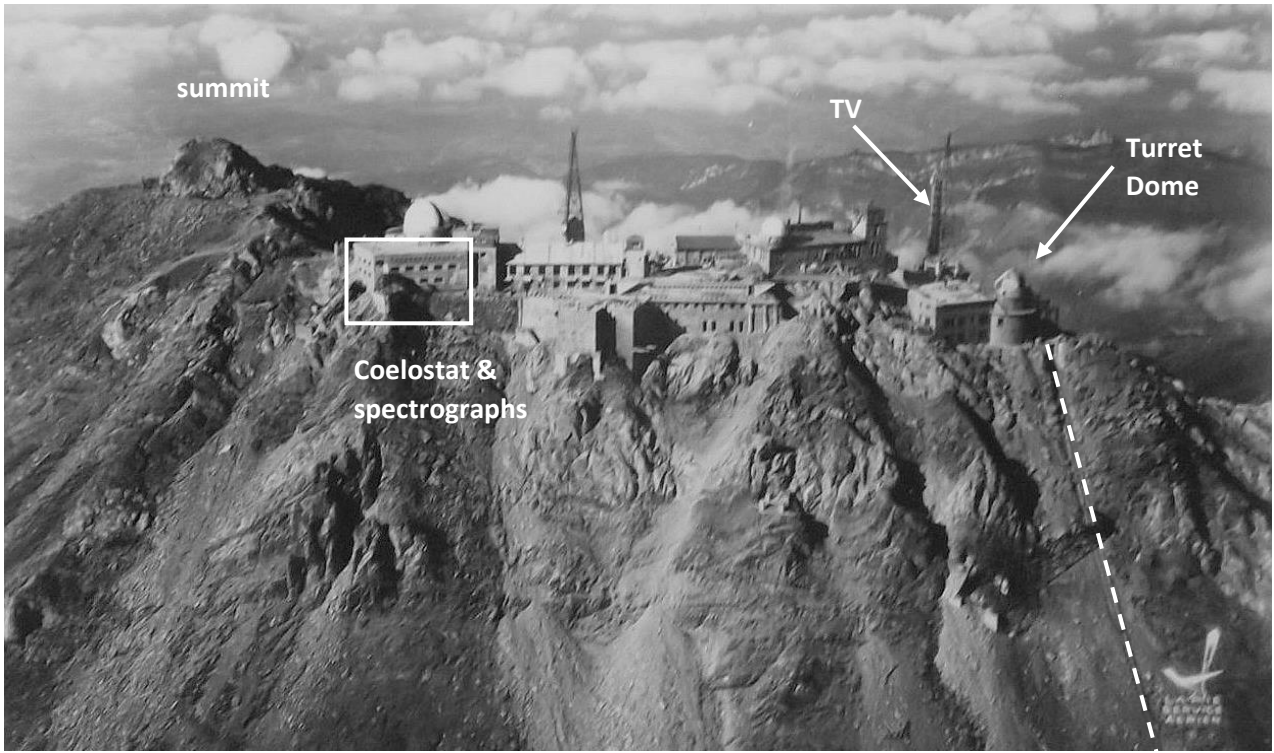

*Figure 8: after 1957. Construction of the solar spectroscopy building under the Baillaud dome in 1956. Television antenna in 1957 (500 W). Turret dome in preparation on the East Tower. Postcard.*

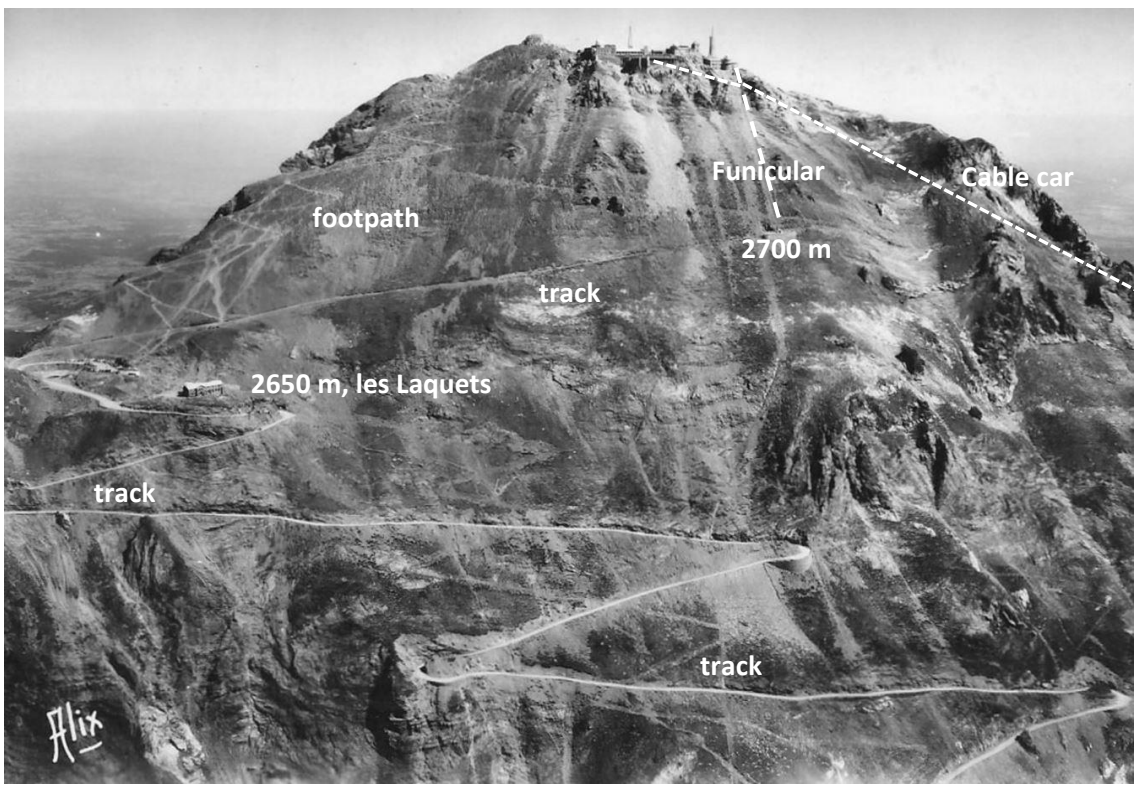

*Figure 9: around 1957. Access. Carriage track 1930. Funicular 1949. Cable car 1952. Postcard.*

The accesses to the Pic were at that time those in Figure 9. After 1957, the clearly visible void on the south side of the terrace between the workshop and the "pas de case" (figure 8) was filled in by the Labardens building (figures 10 and 11) composed of 3 levels of rooms for staff and visitors. Between 1959 and 1962, the summit of the Pic was levelled for the installation of the interministerial building and the large television antenna whose 10 KW transmitter covers 1/7 of the French territory.

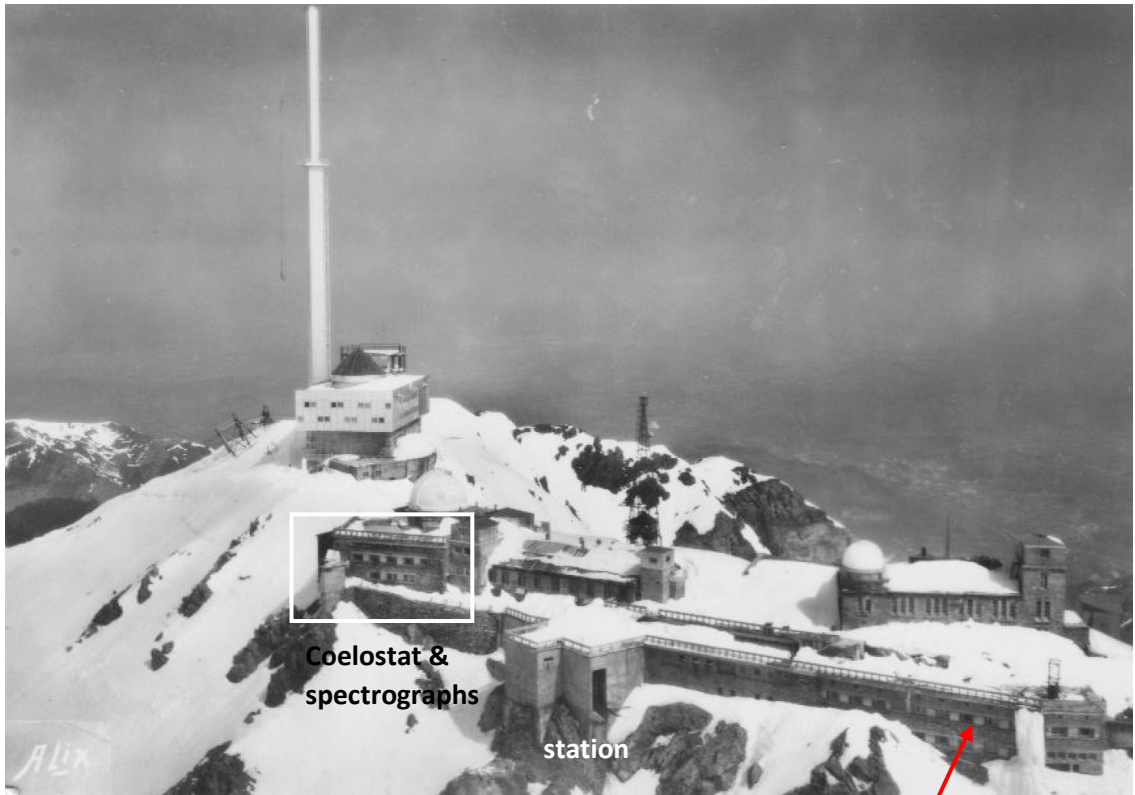

Figure 10: around 1961. The spectroscopy Laboratory in the East has been in operation since 1957. The new television transmitter (10 KW) is being completed. The Labardens building (accomodation, arrow) between the workshop to the west and the "pas de case" to the east is finished. Postcard.

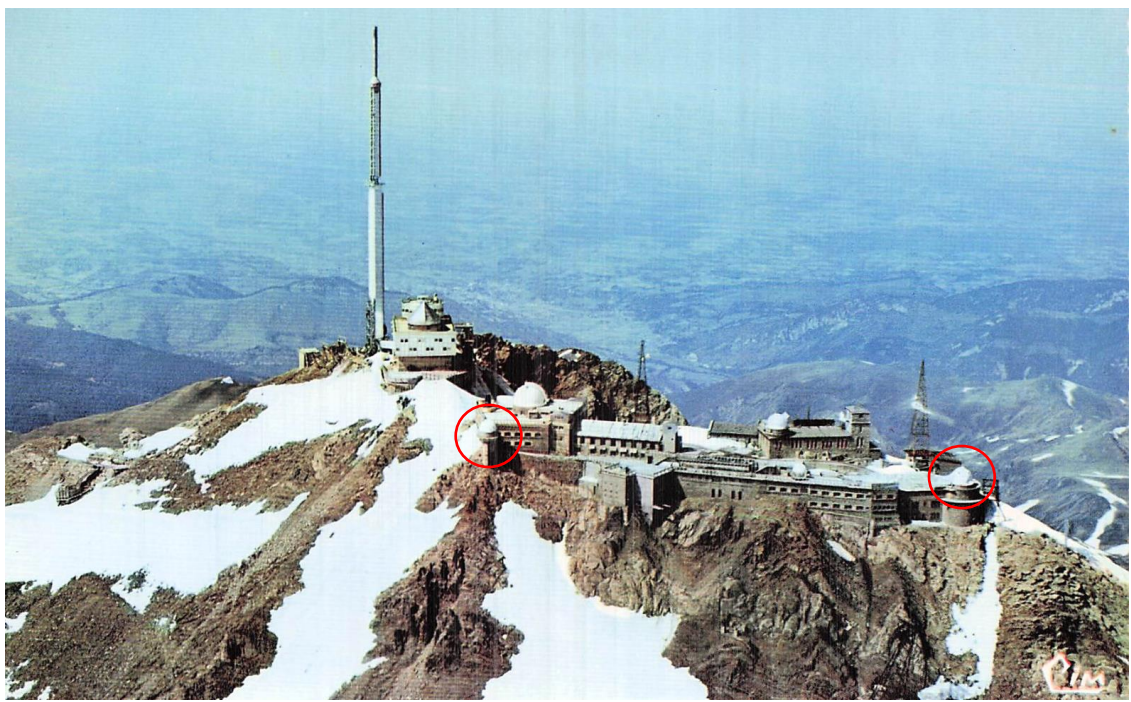

Figure 11: around 1965. New coelostat to the west and turret dome to the east since 1961. Postcard.

In conclusion, the infrastructures of the Pic du Midi in just 14 years from 1947 to 1961 made a prodigious step forward. The entire southern side of the terrace was built, including from west to east, the spectroscopy laboratory (1956), the cable car station (1950), the workshop, the Labardens buildings and the "pas de case" for accomodation, and the east tower supporting the turret dome (1961). As far as access is concerned, two essential advances were produced : the funicular on the east side (1949) and the cable car (1952). The commissioning of the electric power line in 1949 was a major step also. The arrival of the Cosmicians with their Wilson chambers played a driving role in these developments, which were critical for them. On the astronomy side, several new installations were created, for diurnal (solar surface) and night observations (the 60 cm initial telescope of the Gentili dome increased to 106 cm in 1963).

**II – THE SPECTROSCOPY LABORATORY ON THE EAST SIDE OF THE PIC DU MIDI (1956)**

Raymond Michard, head of the solar team at the Meudon Observatory (see Mein & Mein, 2020), proposed an observation program for the International Geophysical Year of 1957/58. The aim was, on the one hand, to intensify the cinematography of chromospheric flares with telescopes equipped with a Lyot filter centred on the Hα line (Meudon and Haute Provence); and on the other hand, to start the study of flares by their line spectrum, in terms of dynamics (Doppler effect) or thermal (multi-line aspect). The concern for high spatial resolution (to do better than in Meudon) then imposed a site like the Pic du Midi where the atmosphere is more stable.

**II – 1 – the coelostat**

The formula adopted for the optical installation was inspired by that of the Meudon spectroheliograph, i.e. a fixed instrument (because of its imposing size) fed by a coelostat with two moving mirrors (Figures 12 and 13). The 50 cm motorized primary compensates for the rotation of the earth (its axis of rotation is North/South), and the secondary, whose position adapts to the seasons, is used to reflect the beam in a fixed direction of the laboratory. It entered by crossing a glass, avoiding convective transfers between the inside and the outside.

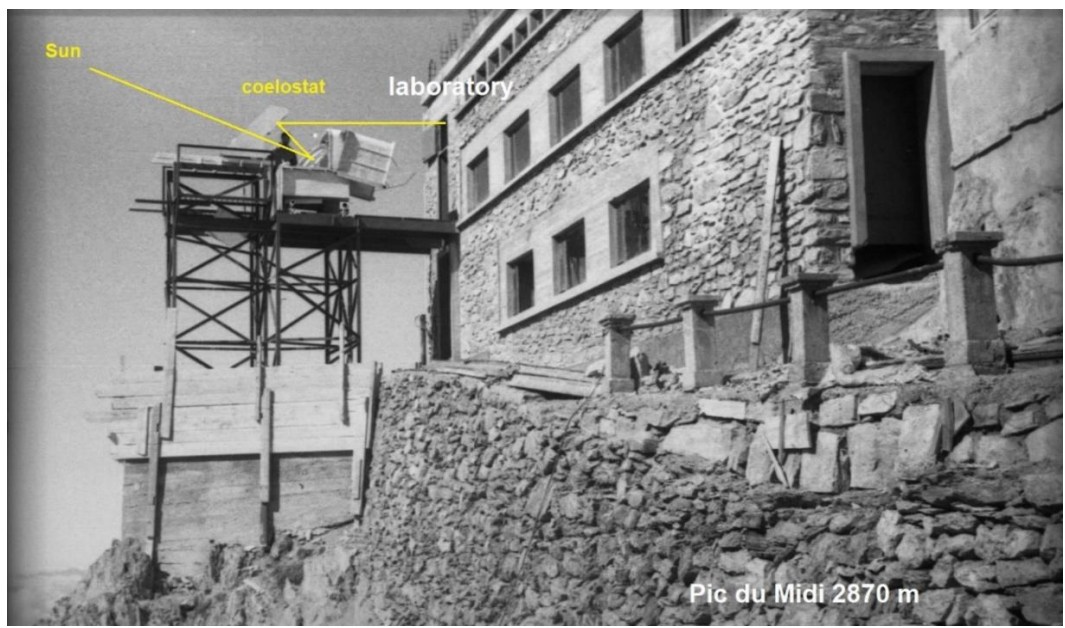

*Figure 12: A preliminary coelostat with two mirrors (40 cm and 50 cm) catched the sun for the horizontal telescope and spectrographs of the Michard laboratory, located west of the Pic (credit OP).*

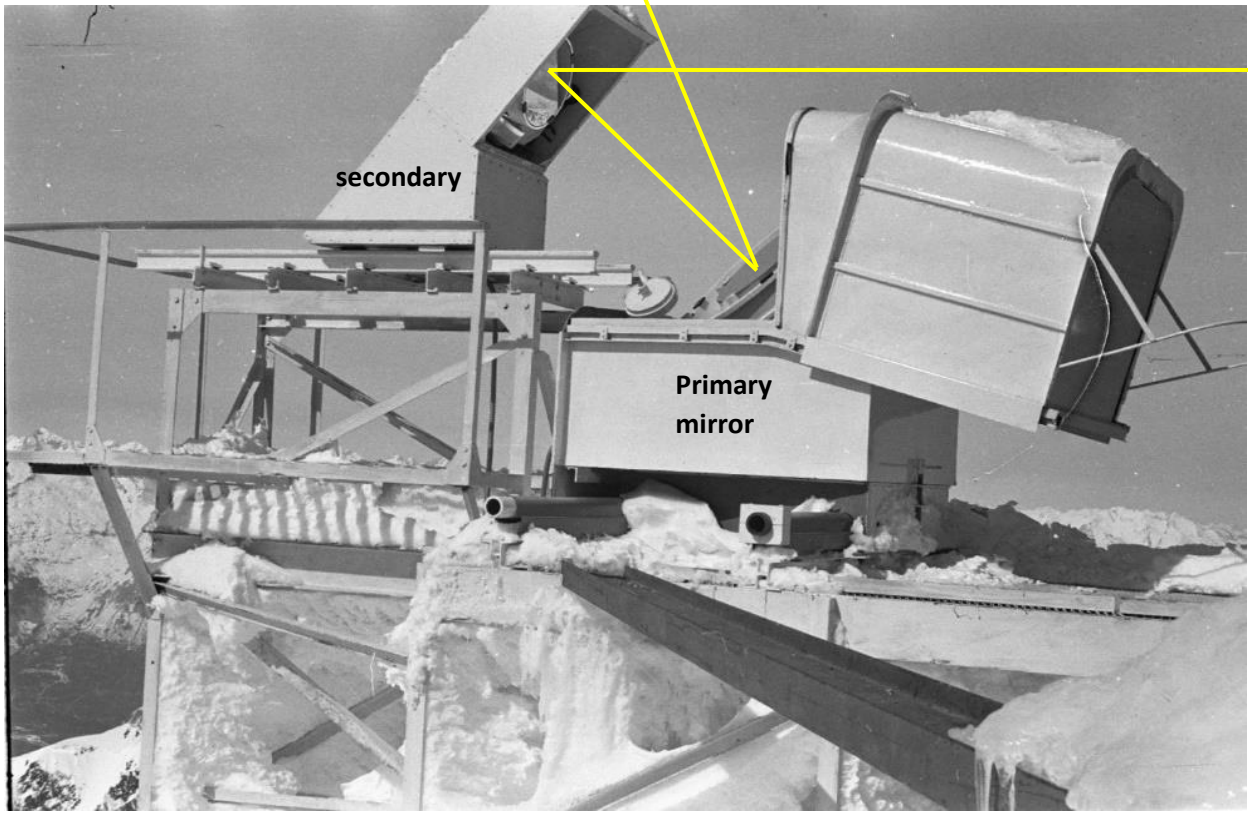

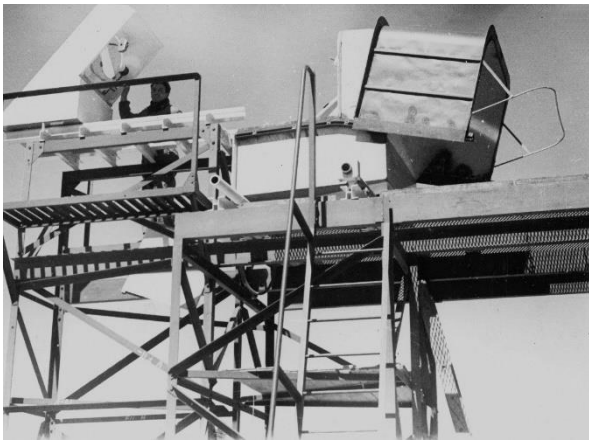
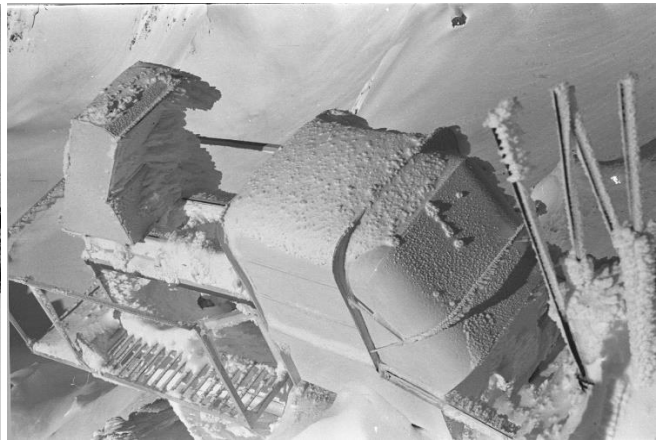

*Figure 13: the first coelostat only had rudimentary protections for its two rail-mounted supports, it was often subjected to the presence of frost and ice at this altitude (credit OP).*

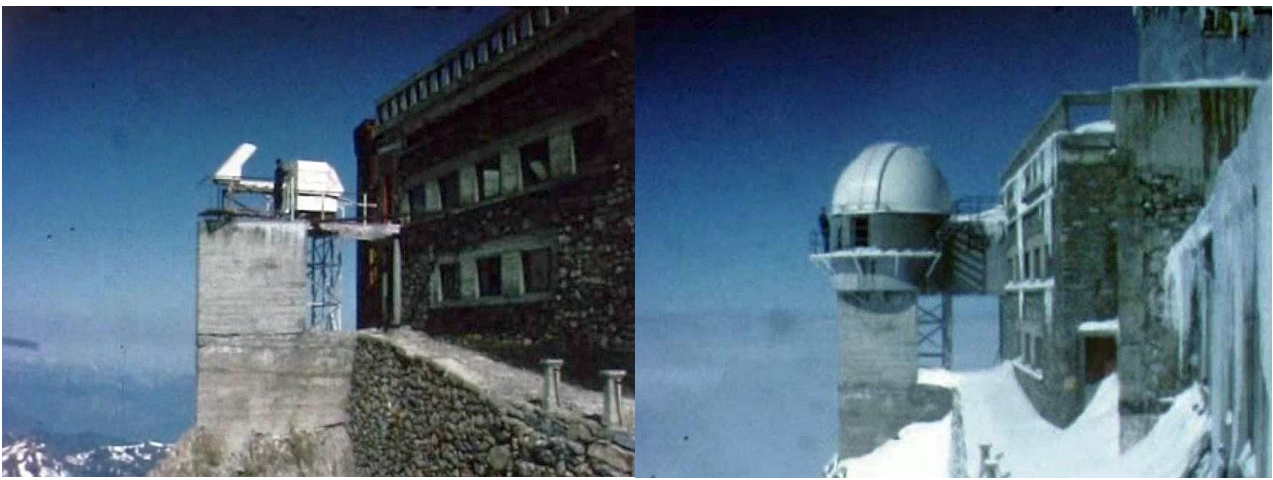

*Figure 14: transition from the first coelostat to the second one under the dome, around 1965 (credit OP).*

The first coelostat, often frozen, was replaced in the 60s by a second under a small protective dome (figures 14 and 15), but it took 3 flat mirrors (the primary and two secondary) to adapt to the small space available under the dome.

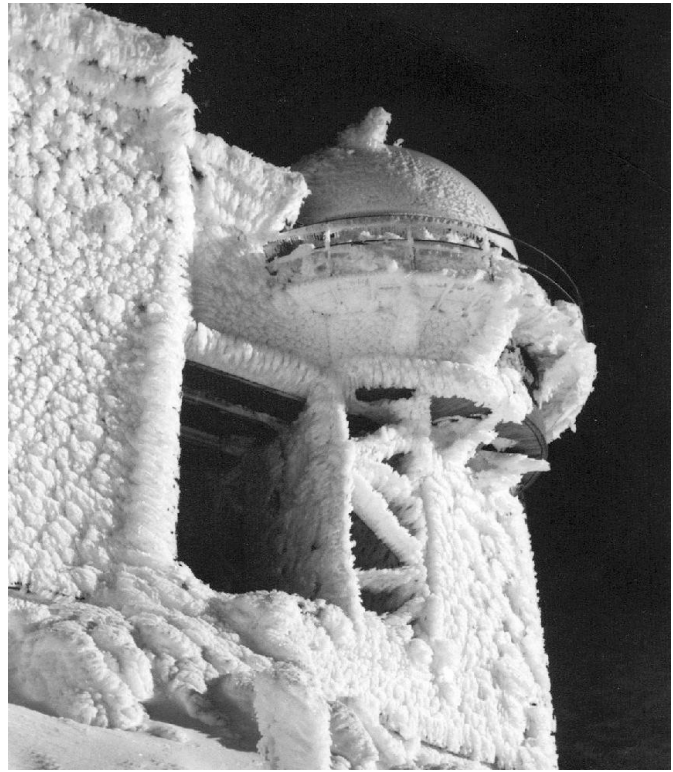

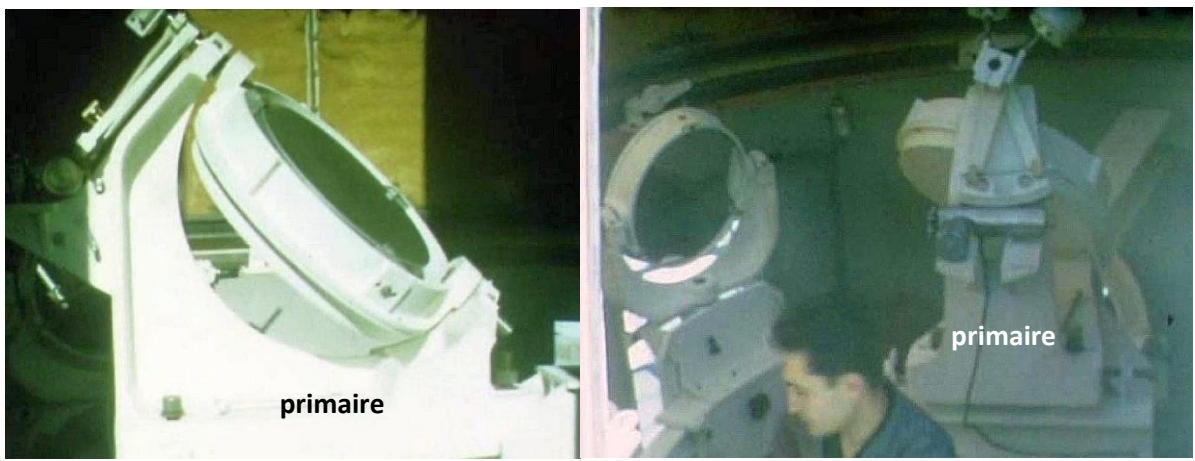

*Figure 15: the second coelostat with 3 flat mirrors under a dome around 1965 (credit OP).*

**II – 2 – the horizontal telescope**

The telescope, fed by the coelostat, was fixed and consisted of a concave mirror 50 cm in diameter and 11 m focal length, giving a 10 cm diameter image on the reflective slit of the spectrographs. This telescope was sent by the Paris Observatory to the Maison du Soleil in Saint Véran, following the full destruction of the laboratory by the Pic 2000 project in order to build the new tourist cable car station instead. A slit jaw device equipped with a Lyot filter, such as those made in series by the company "Optique de Précision Levallois" (OPL), provided an Hα image on a television screen using a video camera, making it possible to point out the structures of solar activity (active centers made of spots, faculae and filaments, potentially eruptive).

## II – 3 – the horizontal 4 m spectrograph

It was a 4 m focal length cross-dispersion spectrograph with prism and grating (Figure 16) providing 3 orders simultaneously, order 4 in UV, 3 in visible and 2 in IR. The collimator was a 4 m concave mirror, while the camera was a 4 m lens objective (Figure 17).

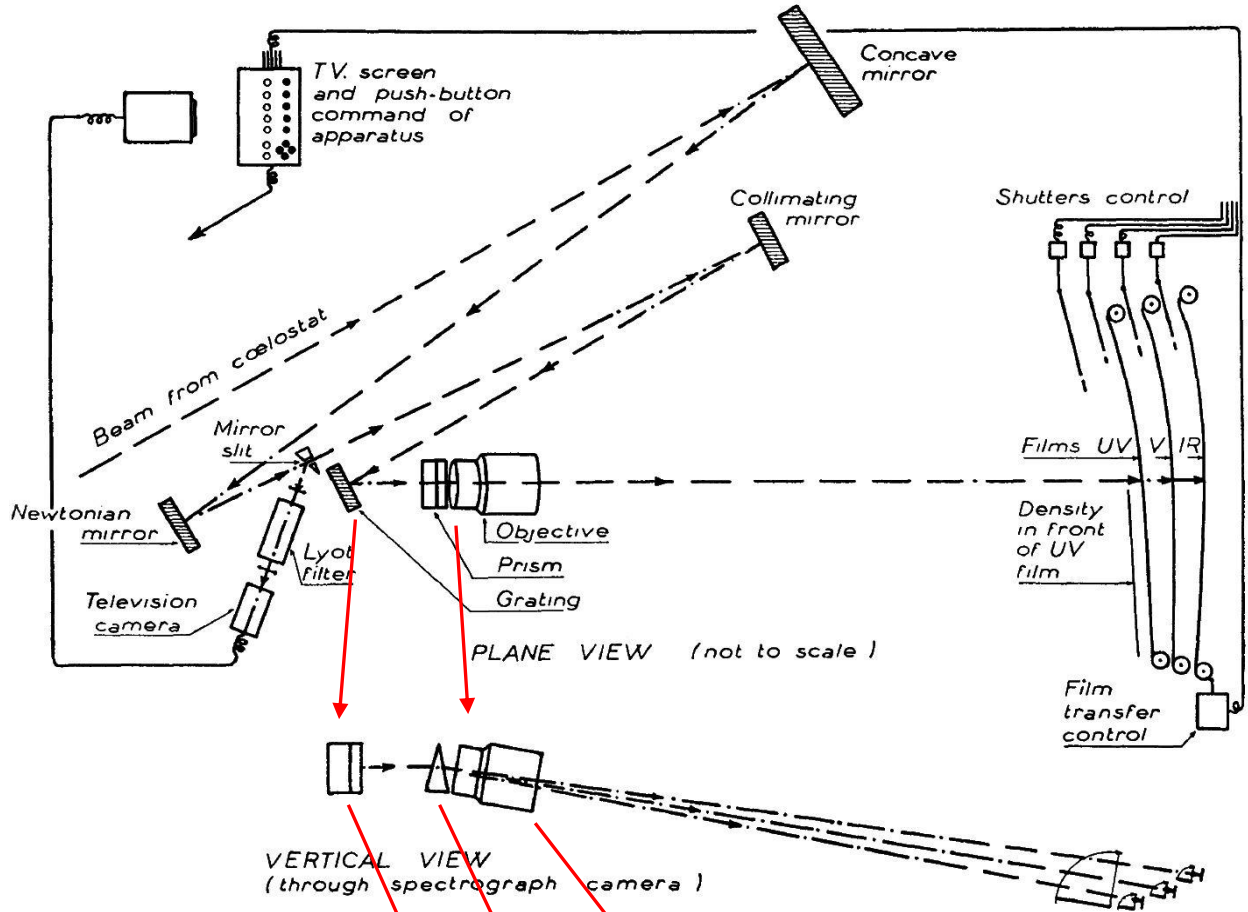

*Figure 16: the 4 m spectrograph of the Michard laboratory (credit OP).*

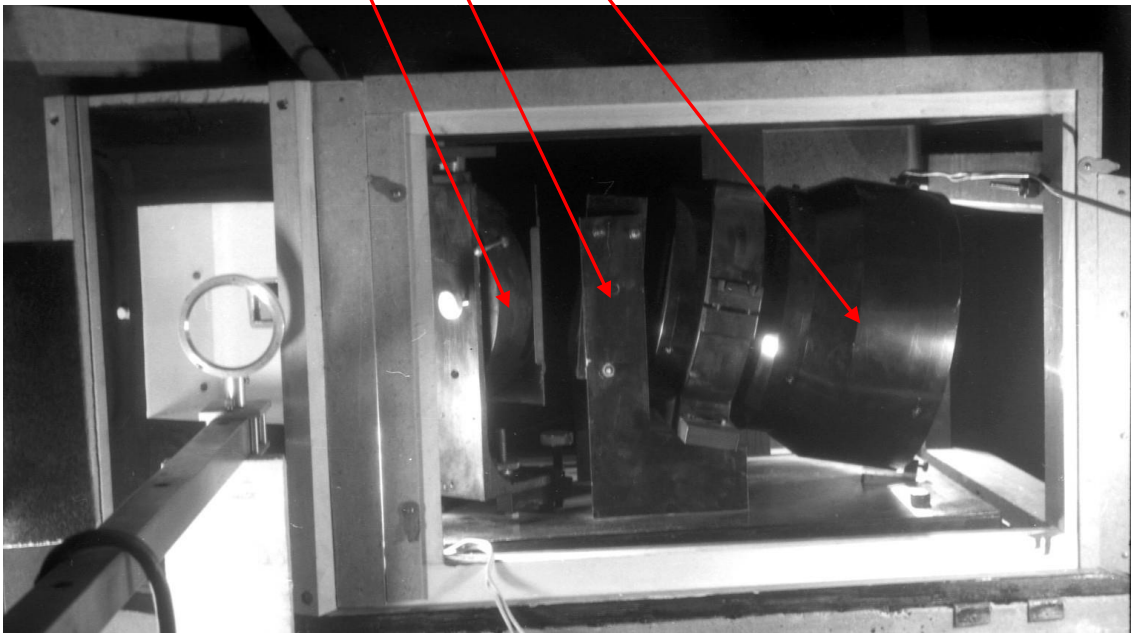

*Figure 17: the grating, the prism for order splitting and the chamber objective (OP credit).*

The dispersive elements consisted of a diffraction grating (300 grooves/mm, 17.5° blaze angle) and a 30° prism splitting orders 2 (IR), 3 (visible) and 4 (UV) focused on three 16 mm films, which were bent by slides to fit the shape of the 3 spectra in space. Figure 18 shows the box containing these 3 horizontal films of about 1 m in length. The advance of the films was motorized. The big advantage of this system, compared to more dispersive spectrographs, was the ability to record the entire visible spectrum and part of the UV and IR spectra in a single exposure, for any point of the Sun located on the 10 mm input slit. The spatial field covered by the slit was of the order of 3 minutes of degree. The dispersion was around 0.3 mm, 0.5 mm and 0.65 mm by Angström, in orders 2 (IR), 3 (visible) and 4 (UV) respectively. These relatively moderate dispersions to study the line profiles in detail subsequently led to the start of the 9 m spectrograph project which we will now discuss. The instrument was controlled by the electro-mechanical programmer console in Figure 19.

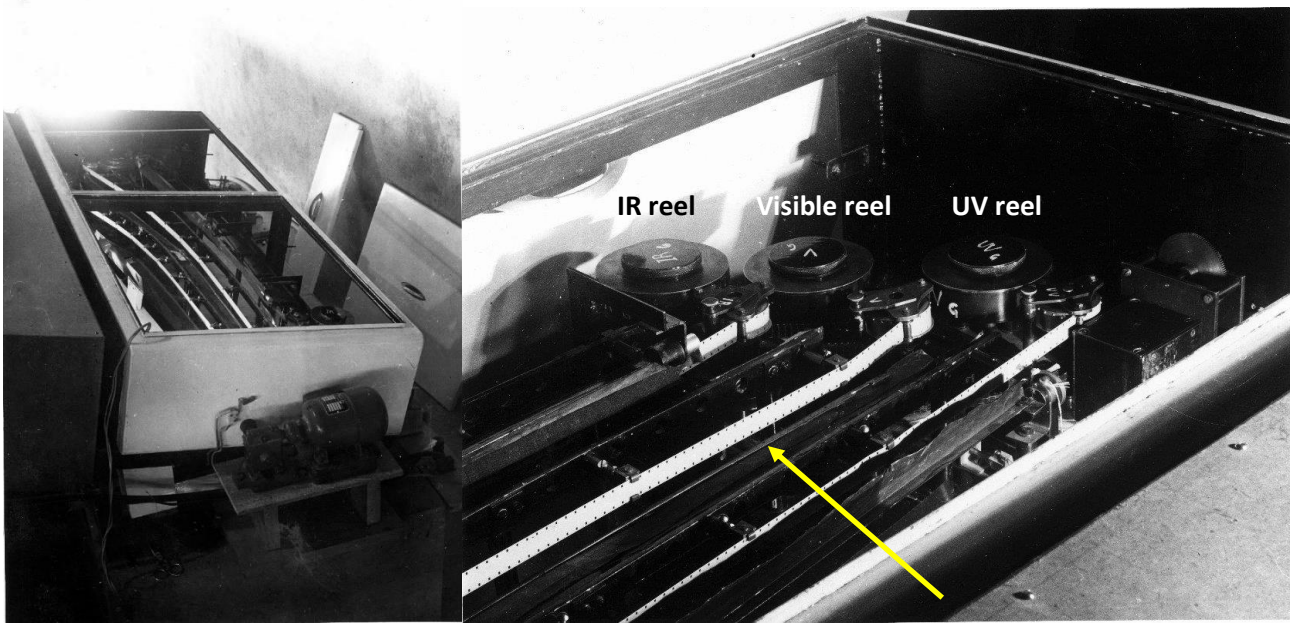

*Figure 18: the 3-film 16 mm box in UV, visible and IR. The reels were motorized (credit OP).*

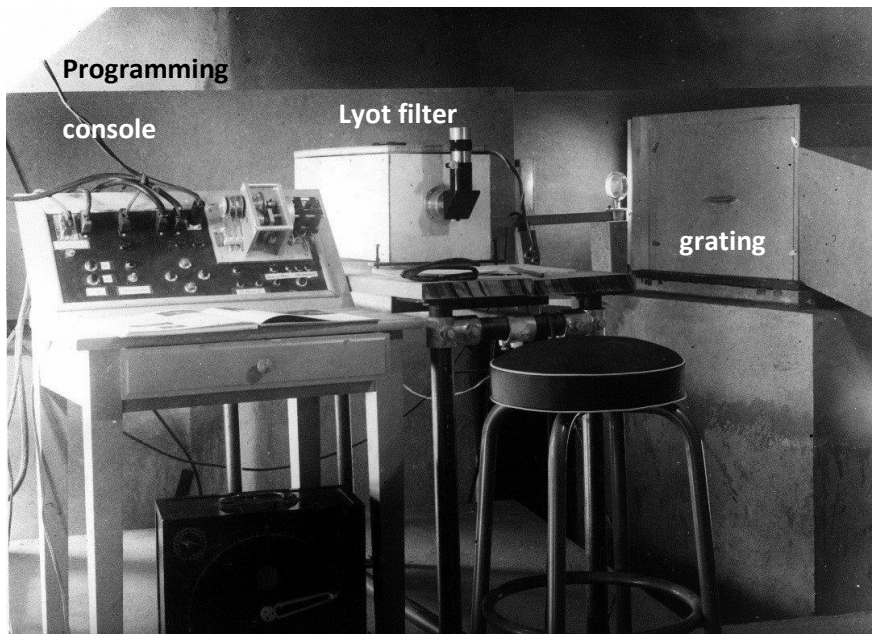

*Figure 19: The 4 m spectrograph control console (credit OP).*

## II – 4 – the 9 m spectrograph

In 1959 the new 9 m spectrograph was added to the laboratory (Figure 20). Designed in Meudon, it was also tested there before being transported to the Pic du Midi. It used the same coelostat and telescope as the 4 m spectrograph, however the two were not usable simultaneously. The 9 m spectrograph included a focal doubler (a concave secondary mirror) providing a focal length equivalent to 22 m, allowing to improve the spatial sampling in case of good images. Both the collimator and the chamber were concave mirrors with a focal length of 9 m (Figure 20). The 600 grooves/mm grating, blaze angle 55°, provided portions of the visible spectrum in orders 3 to 8 with an average dispersion of 4 mm/Angström and a resolution of 0.01 Angström perfect for the detailed study of the thin lines of the photosphere (a big improvement over the 4 m spectrograph which was limited to 0.07 Angstrom).

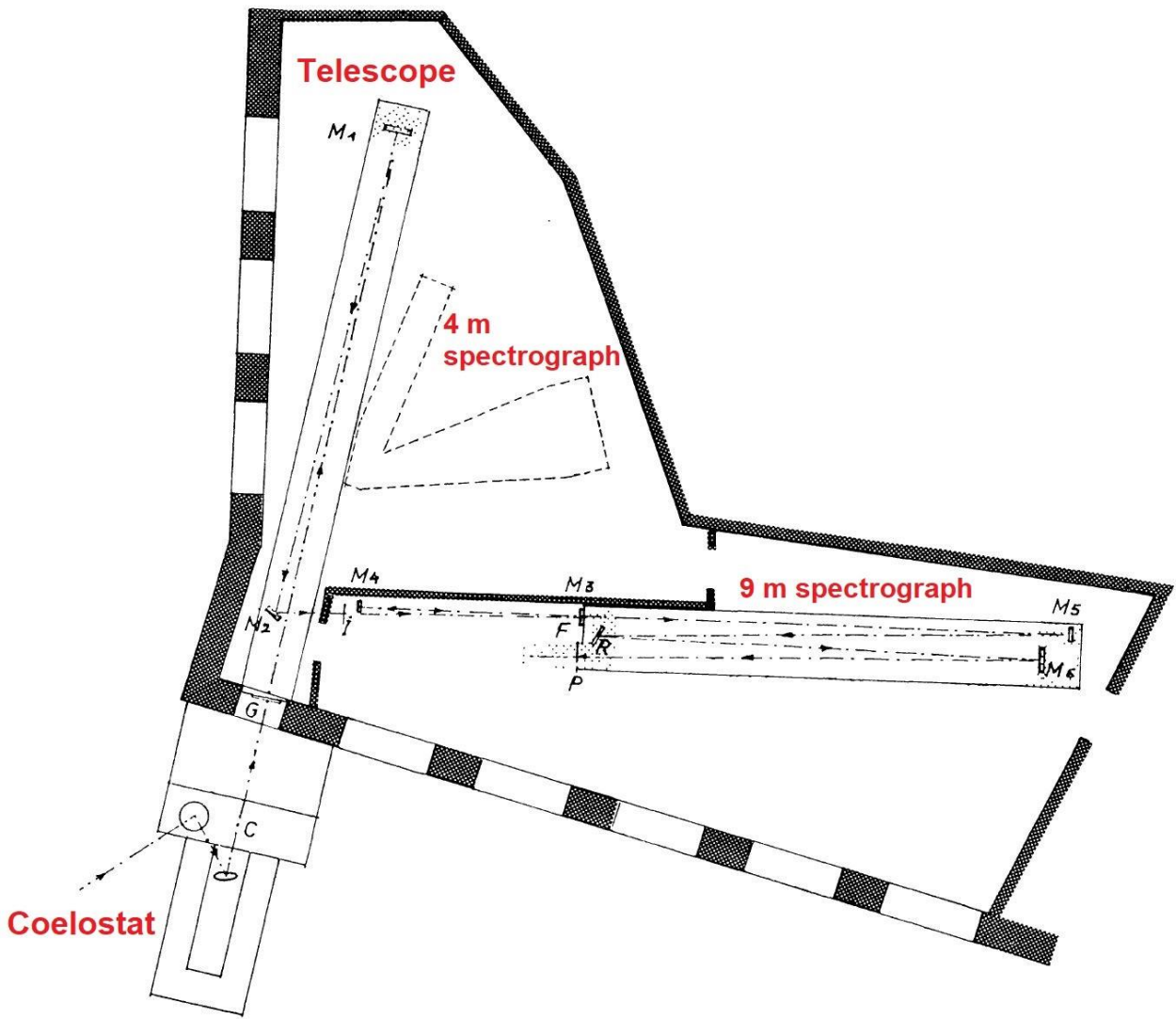

C : coelostat
G : uviol glass
M1 : 11 m telescope
I : primary image (dia 10 cm)
M2 : flat mirror
M3 : magnifier mirror (x 2)
M4 : flat mirror
F : reflecting slit (20 cm solar diameter)
M5 : 9 m collimator
R : 600 grooves/mm grating (20°/55° blaze)
M6 : 9 m chamber
P : spectrum (20 cm solar image height)

*Figure 20: the 9 m spectrograph (credit OP).*

At the focus of the spectrograph, it was possible to use 9 x 24 cm² photo plates or 35 mm films for fast shooting. In Figure 21, we see Roger Servajean, a Michard's collaborator, at the exit of the spectrograph, and behind him the control console. The diffraction grating is shown in Figure 22.

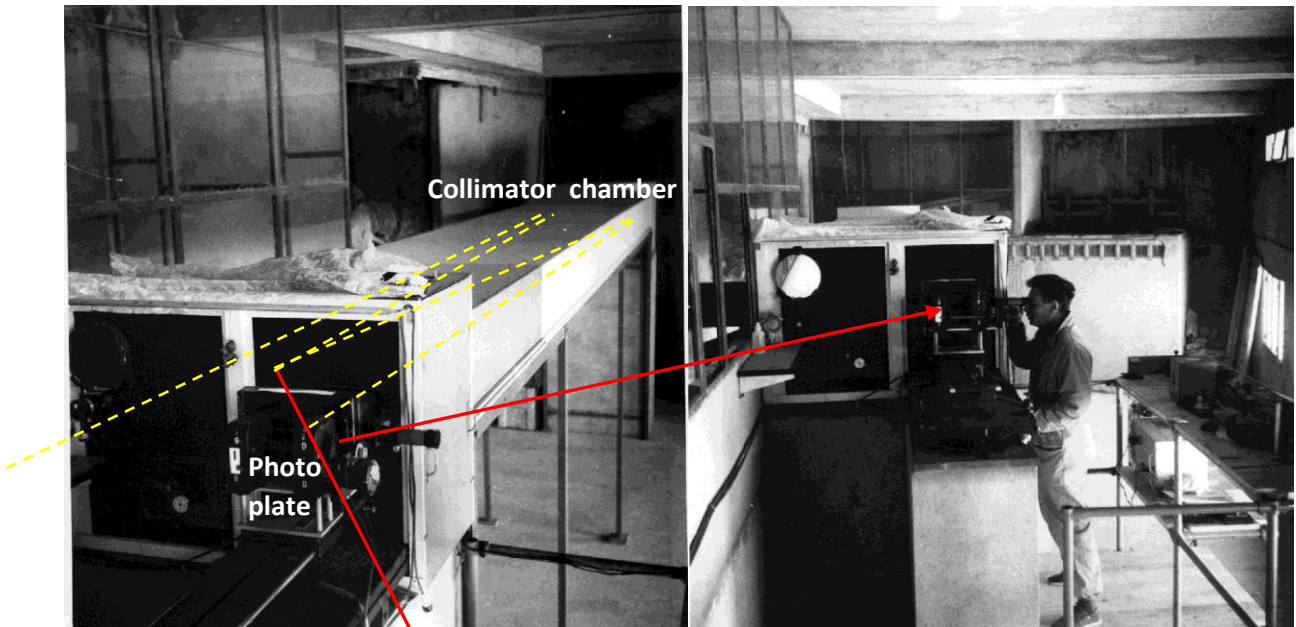

*Figure 21: The inlet (left compartment) and the outlet (right compartment) of the 9 m spectrograph. The visible mirror is the focal enlarger x 2 (from 11 to 22 m, credit OP).*

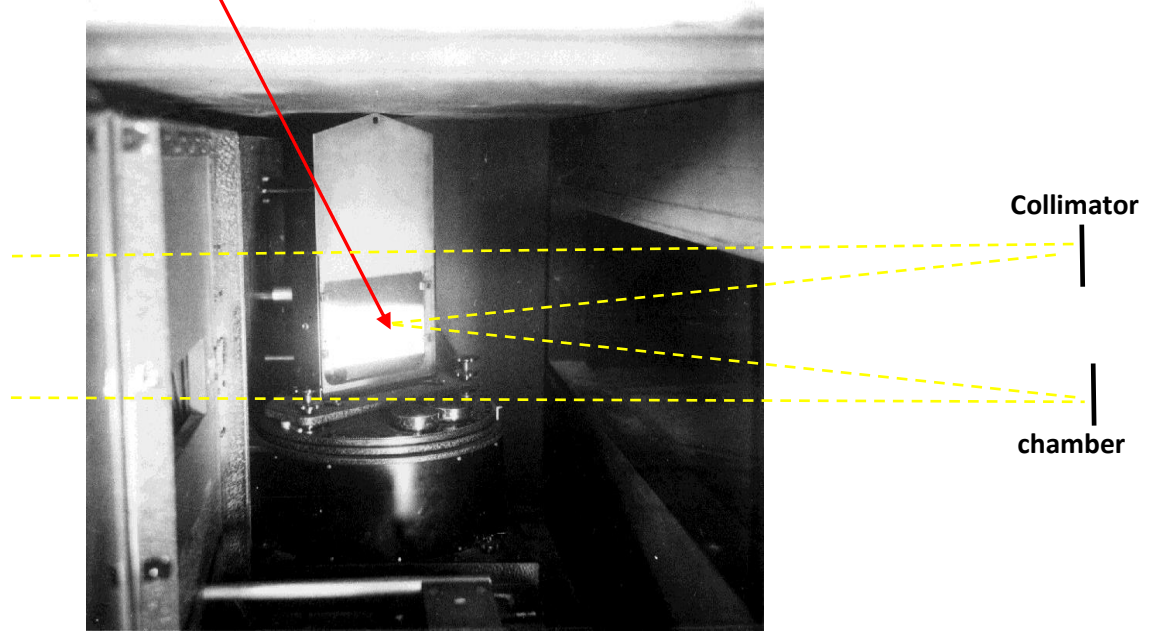

*Figure 22: the diffraction grating of the 9 m spectrograph (credit OP).*

The spectrograph's entrance slit, reflective, was equipped with a sophisticated imaging system to assess the active structures to be observed. There were:

- a Lyot Hα filter with film camera and eyepiece, for the chromosphere

- a white light slit jaw with film camera, Leica reflex camera, projection screen and eyepiece, for the photosphere

A Hale grid was put in front of the slit in case of Zeeman component analysis (Figure 23).

Figure 23 below displays an example of a solar flare observed with the 4 m spectrograph, while Figure 24 shows the Zeeman effect on a spectrum observed through a Hale grid with the 9 m spectrograph (alternating quarter- and three-quarter wave strips plus linear polarizer).

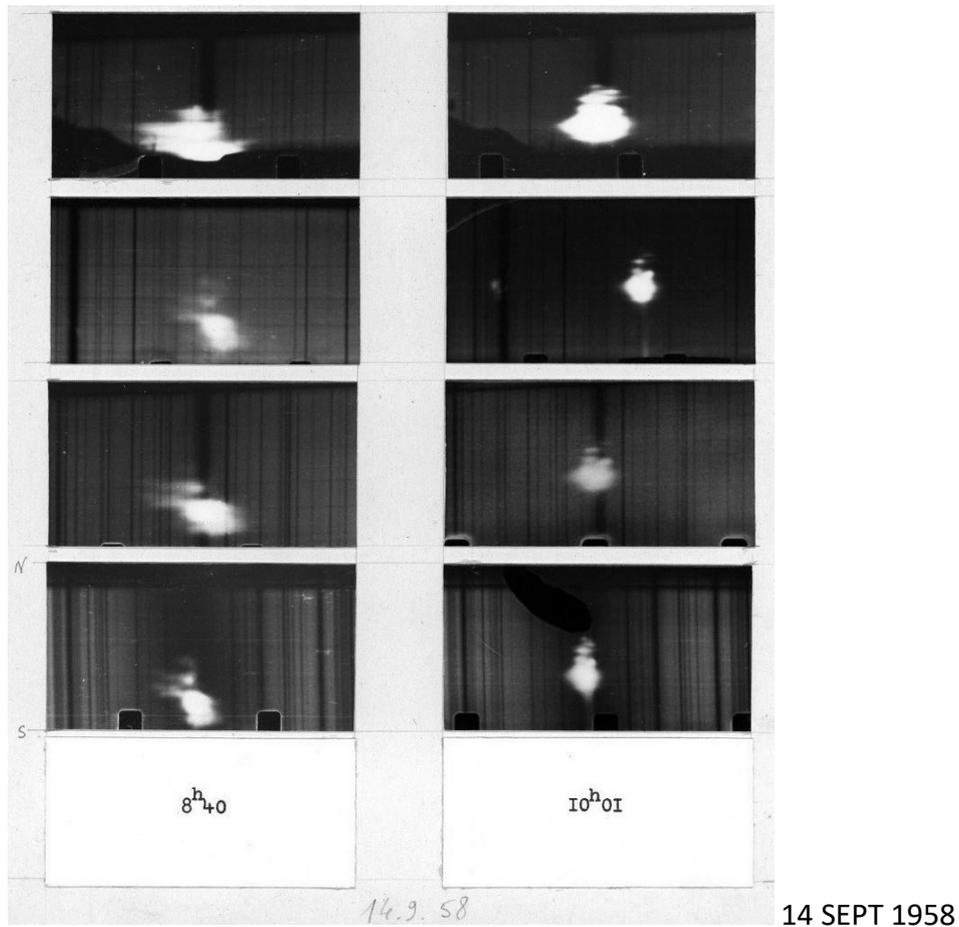

14 SEPT 1958

*Figure 23: Two times of a flare in 4 spectral domains with the 4 m spectrograph (credit OP).*

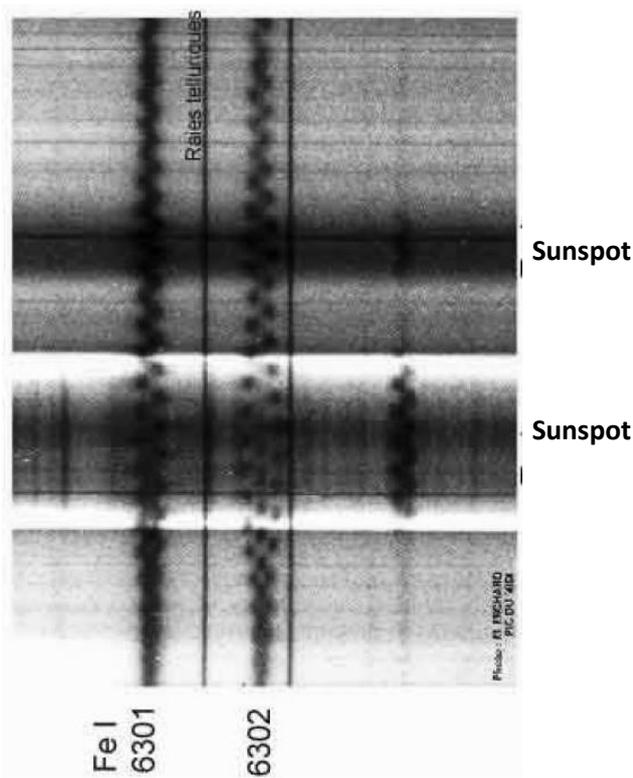

*Figure 24 :*

*Spectrum of the FeI 6301.5 and FeI 6302.5 Angström lines in an active region with sunspots obtained with the 9 m spectrograph by Michard and his team (see Michard et al, 1961).*

*The spectrum is modulated by a Hale grid which alternately transmits the σ+ and σ- components of the Zeeman effect, causing a zigzag of the line that is more pronounced when the Lande factor is high (1.67 for 6301.5, 2.5 for 6302.5). The σ+ and σ- measurements provide together the line of sight magnetic field, but they are not cospatial with this method, providing rough results (this is only of historical interest today in solar polarimetry). There are 2 telluric lines in the spectral field. Credit OP.*

## III – THE TURRET SPECTROGRAPH AT THE EAST CREST OF THE PIC DU MIDI

The turret dome is the work of Rösch (Figure 25). Built in Bagnères before 1960, it was mounted to the Pic by the funicular and put into service with a 38 cm refractor (Figure 25).

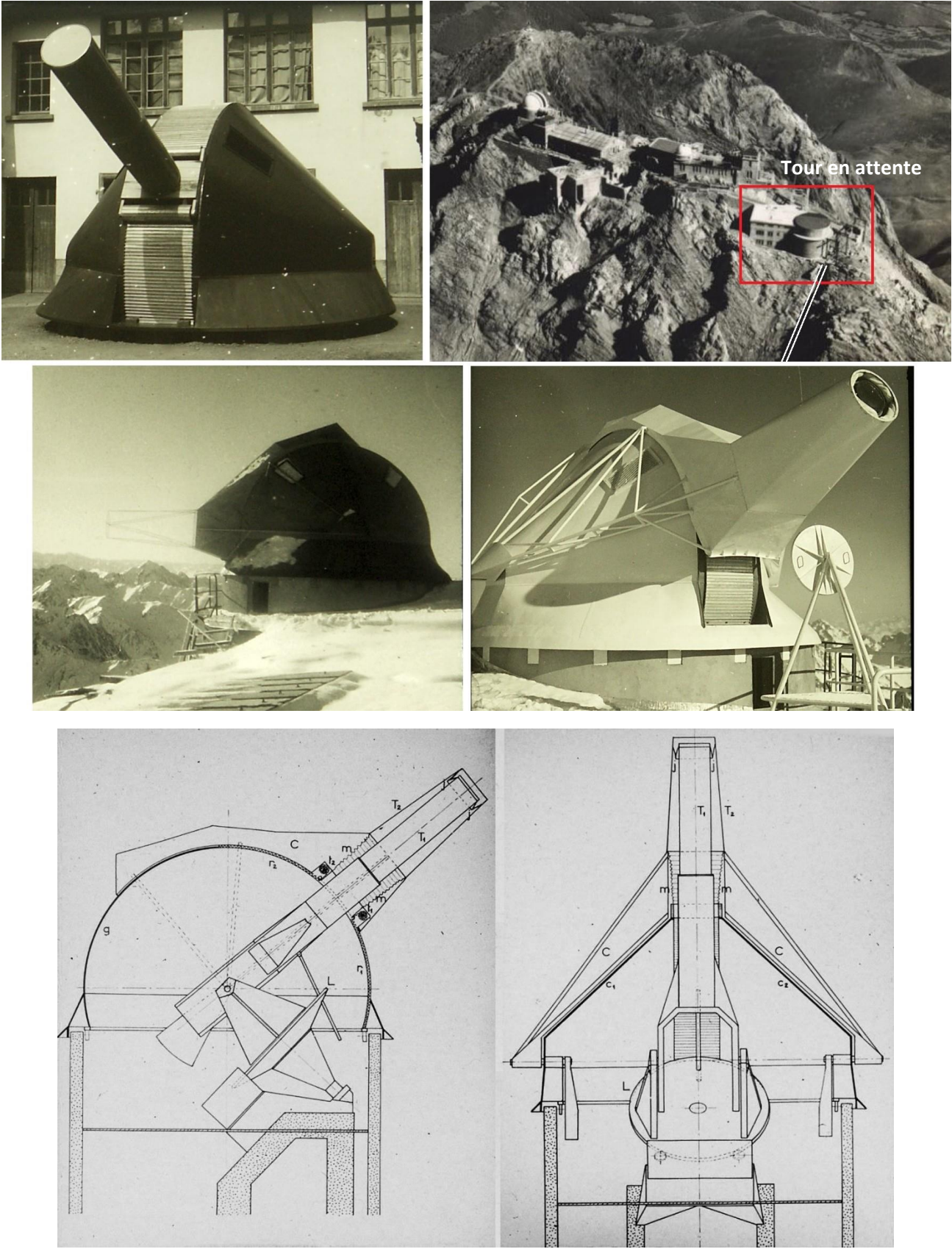

*Figure 25: the turret dome with its lens rejected out of the closed cupola (credit OMP).*

The aim of this project (Roudier *et al*, 2022) was to build an instrument able of delivering the best image quality allowed by the site (without active optics which did not exist at that time). The best location was the eastern crest of the Pic in the direction of the rising sun. The lens was pushed away from the cupola in order to avoid its turbulence, and the dome remained closed by movable metallic curtains to prevent any exchange of air between the inside and the outside. In 1975, the 38 cm lens was exchanged for a more optically optimised 50 cm refractor. This instrument, operated by Richard Muller and Thierry Roudier, produced images of unprecedented quality, which is why it was used in a similar form (the dome with a circular hole) on the 2 m night telescope in 1980 (the TBL) and then on the 90 cm THEMIS solar telescope in the Canary Islands after 2000. The two-glass lens had a focal length of 6.5 m that was increased to 15, 30 or 60 m using an enlarger. Around 1980, Mouradian (from Meudon), a regular user of the 9 m spectrograph, decided to mount on the refractor an 8 m spectrograph with a grating (300 grooves/mm, 63° blaze angle) operating in orders 5 to 15, and providing an average dispersion of 5 mm/Angström (Figure 26).

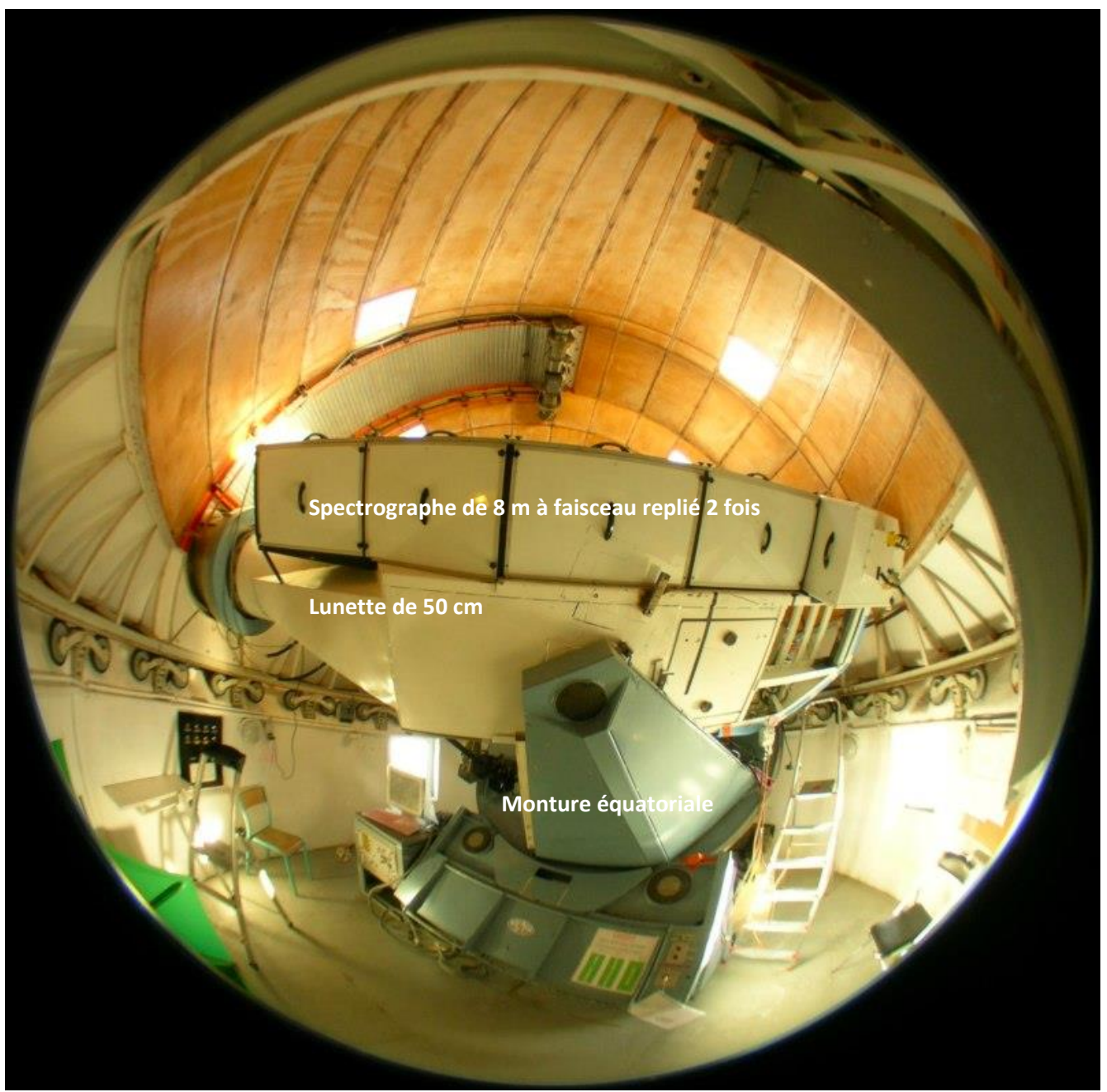

*Figure 26: the refractor mount and turret spectrograph in their closed dome (credit S. Rondi).*

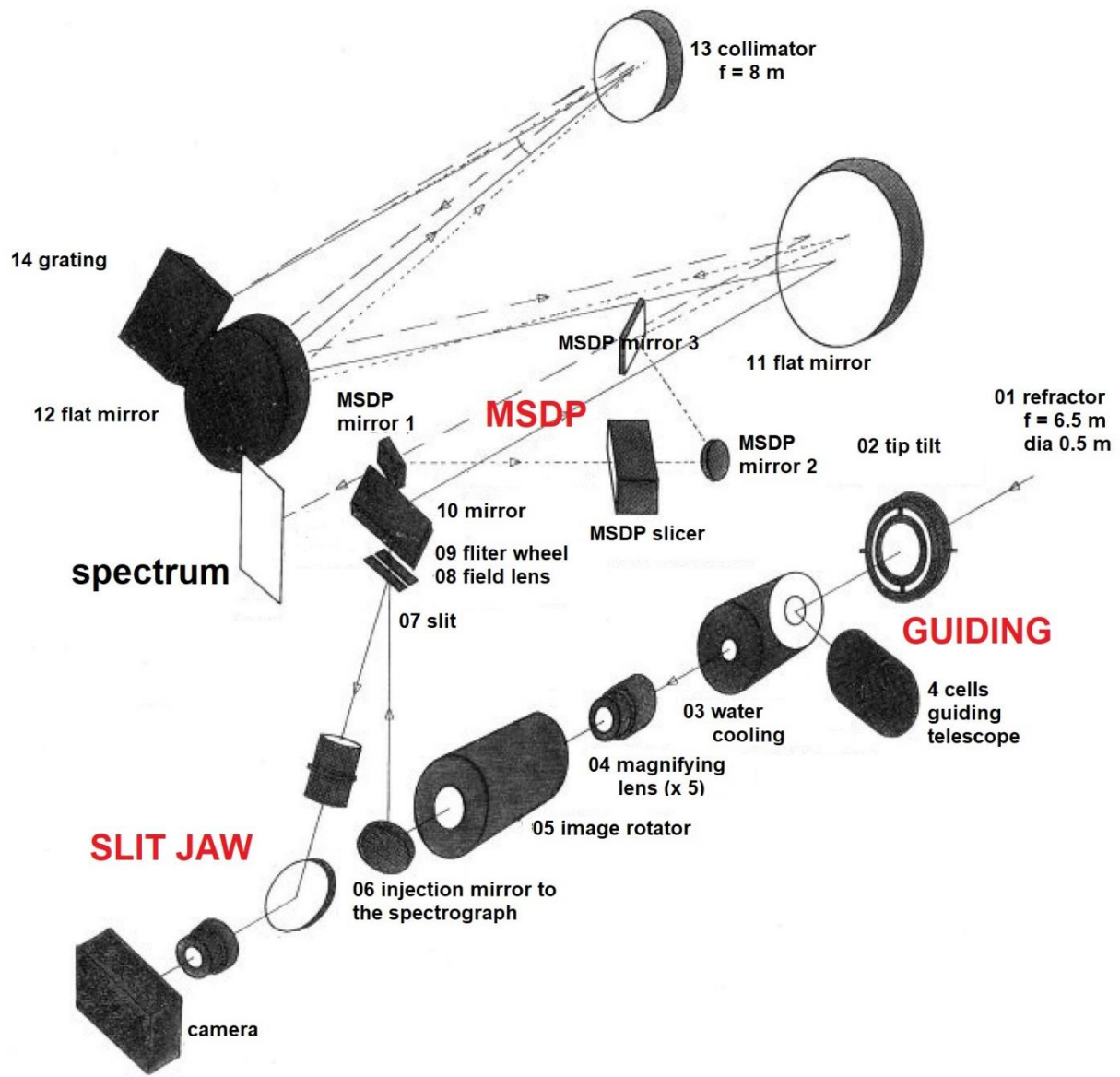

*Figure 27: diagram of the 8 m turret spectrograph (after Mouradian, 1980, credit OP).*

The spectrograph (Figure 27) was of the Littrow type with an 8 m spherical mirror and two planar mirrors bending the beam twice so that the length of the spectrograph did not exceed 3.5 m, because the available space on the equatorial mount of the telescope was reduced (Mouradian, 1980). This instrument, made in Meudon, was very complete : there was a slit jaw, a guiding system based on four photoelectric cells, an optical image rotator and a 2D imaging spectroscopic device, the Multichannel Subtractive Double Pass (MSDP with 11 2D channels) introduced by Pierre Mein (Figure 27 and Mein, 1980). The detector remained the photographic plate (Coleman cameras with 70 mm films, figure 28) until the end of the 90s when a 1280 x 1024 CCD was implanted after a transfer optics had been made, since the spectrum of 6 x 9 cm² had to be reduced 10 times to enter the CCD target. This spectrograph was combined with a fast liquid crystal polarimeter from 2003 (Malherbe *et al*, 2007), taking advantage of the almost absence of parasitic instrumental polarization. From 2006, the turret dome was progressively dethroned in terms of imaging quality by the HINODE space telescope, of the same diameter, and in terms of spectro-polarimetry by more modern instruments such as ZIMPOL at ETH/Zürich or THEMIS. As a consequence, professionals gradually deserted the Pic du Midi, and were replaced by university students for training sessions.

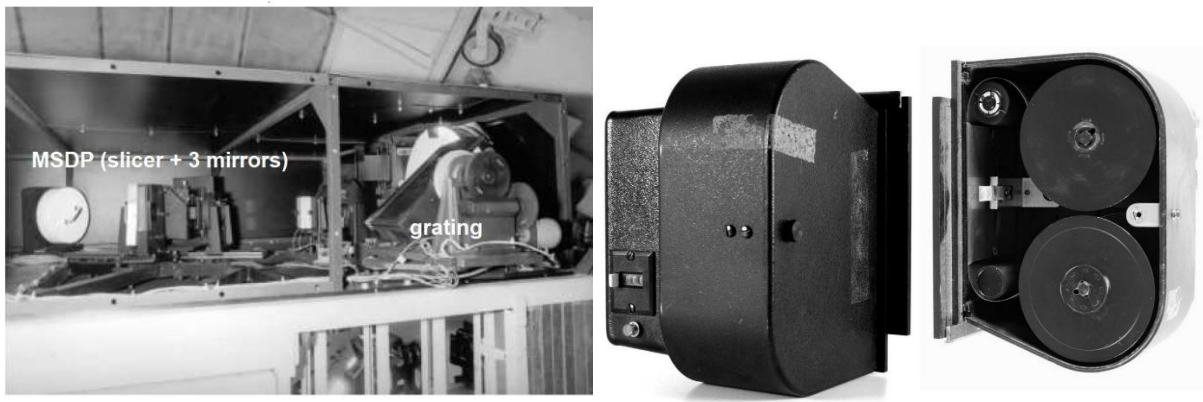

*Figure 28: on the left, the spectrograph with the grating (300 grooves/mm, blaze 63°26') and the MSDP device in the spectrum. On the right, Coleman cameras with 70 mm film reels used in direct output.*

**IV – SPECTRO CORONAGRAPHS**

It should be remembered that coronagraphs are intended to study the corona beyond the limb and hide the bright solar disk. The spectrograph mounted on the Lyot coronagraph (13 cm aperture, 3.15 m focal length) in 1931 (Lyot, 1931, figure 29) had the following characteristics : a concave Rowland grating of 7 m radius (3.5 m focal length) at order 2 providing a dispersion of 0.8 mm/Angström. A second 61° prism device with a focal length of 3.6 m gave a lower dispersion, i.e. 0.3 mm/Angström, but it was 15 times brighter for the green line of Fe XIV. Many emission lines of iron were discovered or observed by Lyot (table 1); the highest ionization state corresponds approximately to the highest temperature, which can vary from $10^6$ K to $10^7$ K in the corona. Many iron lines are present in the EUV spectrum (below 400 Angström), and therefore are visible only from space (SOHO mission, 1996 ; SDO satellite, 2010, Solar Orbiter, 2020). They are emissive and dominate on the solar disk (because the photospheric blackbody at 6000 K mainly emits in the visible, so that a coronagraph is not necessary in EUV wavelengths). But before the space era, ground based spectroscopy with coronagraphs in mountains was the only tool to explore the corona above the limb at various temperatures.

| λ | Fe X | Fe XI | Fe XII | Fe XIII | Fe XIV | Fe XV |
|---|---|---|---|---|---|---|
| 3 010 | . | . | x | . | . | . |
| 3 388 | . | . | . | x | . | . |
| 3 987 | . | x | . | . | . | . |
| 5 303 | . | . | . | . | x | . |
| 6 374 | x | . | . | . | . | . |
| 7 060 | . | . | . | . | . | x |
| 7 892 | . | x | . | . | . | . |
| 10 747 | . | . | . | x | . | . |
| 10 798 | . | . | . | x | . | . |

*Table 1 : lines of ionized iron in the visible, close UV and close IR (after Hugon et al, 1963)*

Hugon *et al* (1963) mention a second Littrow-type Lyot spectrograph with a single concave mirror of one meter focal length (table 2) and a plane diffraction grating of 600 grooves/mm giving a dispersion of 0.13 mm/Angtröm for the green line of Fe XIV (Figure 30) that was mounted on another 20 cm aperture, 4 m focal length Lyot coronagraph. They also cite another spectrograph (n° I of table 2), of the Ebert-Fastie type with a mirror of 52 cm focal length, grating 600 grooves/mm, blaze 17°27', giving 0.07 mm/Angström dispersion, for a third 15 cm aperture coronagraph (1.5 m focal length) as well as two other spectrographs n° II and n° III (table 2) designed for a future instrument of 26 cm aperture. Demarcq *et al* (1965) describe a spectrograph similar to the I, but with the Czerny-Turner formula with two mirrors of 52 cm focal length mounted on the 15 cm coronagraph. Numerous spectro-coronagraphic studies have been carried out by Jacques-Clair Noens (e.g. Noens *et al*, 1984, for measurements of electron densities from Fe XIII in the IR) or by Jean-Pierre Rozelot (1972). The spectrograph described by Demarcq *et al* (1965) was also used by Rozelot & Despiau (1972, figure 31) on the turret dome refractor from which the objective had been removed and replaced by the 15 cm coronagraph. The Lallemand electronic camera was tested as an ultra-sensitive detector, a real first (as it was designed for deep night sky). However, this experiment was not repeated due to the difficult removal and repositioning of the 50 cm lens of the turret dome refractor, since two instruments could not coexist on the mount of this closed cupola. The Lallemand camera was reused later by Paul Felenbok at Saint Véran's coronagraph in 1976.

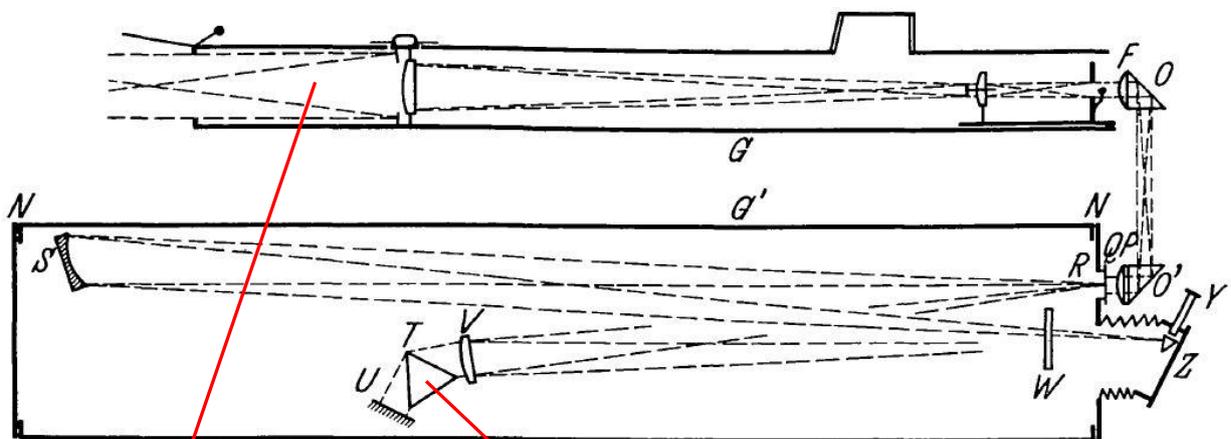

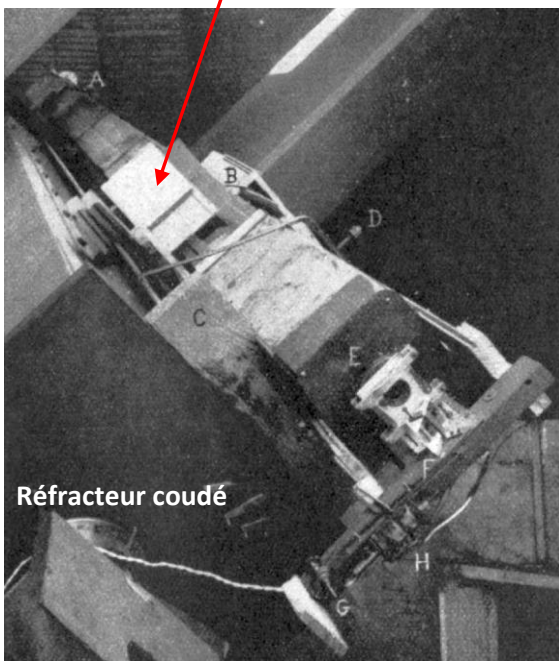

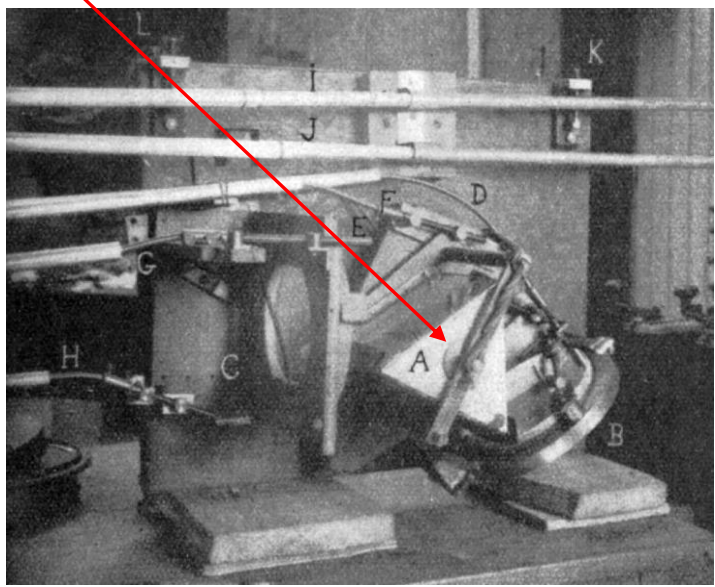

Figure 29: Bernard Lyot's first spectro-coronagraph (diagrams by Lyot, 1931, photos by Lyot, 1932)

| Spectrographe | Lyot | I | II | III |
|---|---|---|---|---|
| Type | Littrow | Ebert-Fastie | d⁰ | d⁰ |
| Nombre de traits par mm | 600 | 600 | 600 | 600 |
| Ordre pour 5303 Å | 2 | 2 | 3 | 5 |
| Pouvoir de résolution | 72 000 | 36 000 | 180 000 | 600 000 |
| Longueur focale cm | 100 | 52 | 150 | 400 |
| Dispersion Å/mm | 7 | 15 | 3 | 0.7 |

Table 2: the spectro coronagraphs of the Pic du Midi (after Hugon et al, 1963)

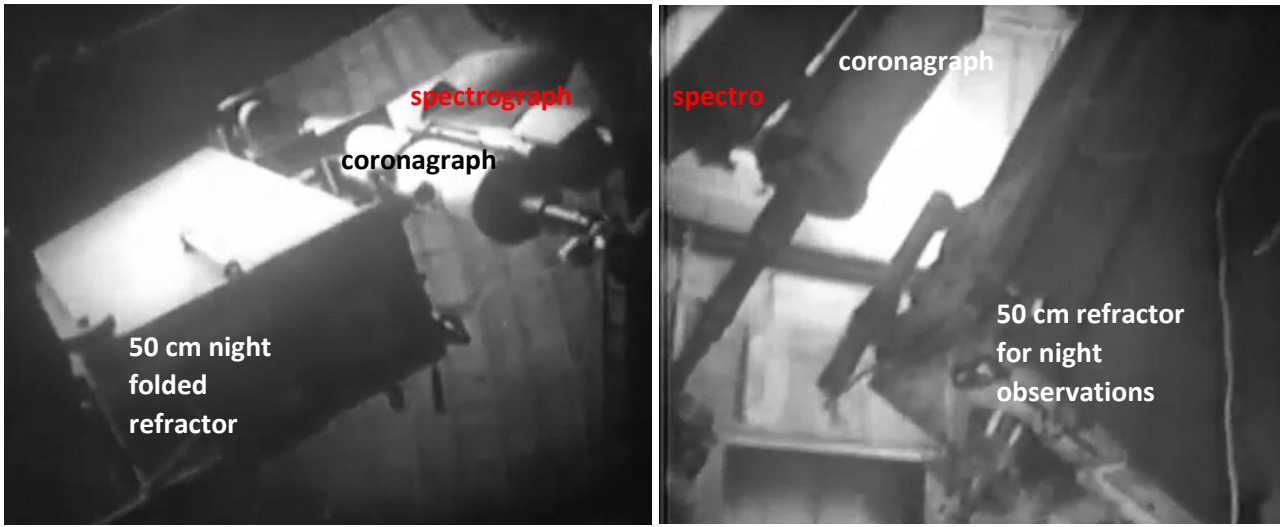

Figure 30: the second Lyot spectro-coronagraph (20 cm aperture) with a shorter spectrograph, focal length of 1 m (after a film by Joseph Leclerc, "The Hermits of the Sky"). Lyot's instruments coexisted on the mount of the Baillaud dome with the folded nocturnal refractor.

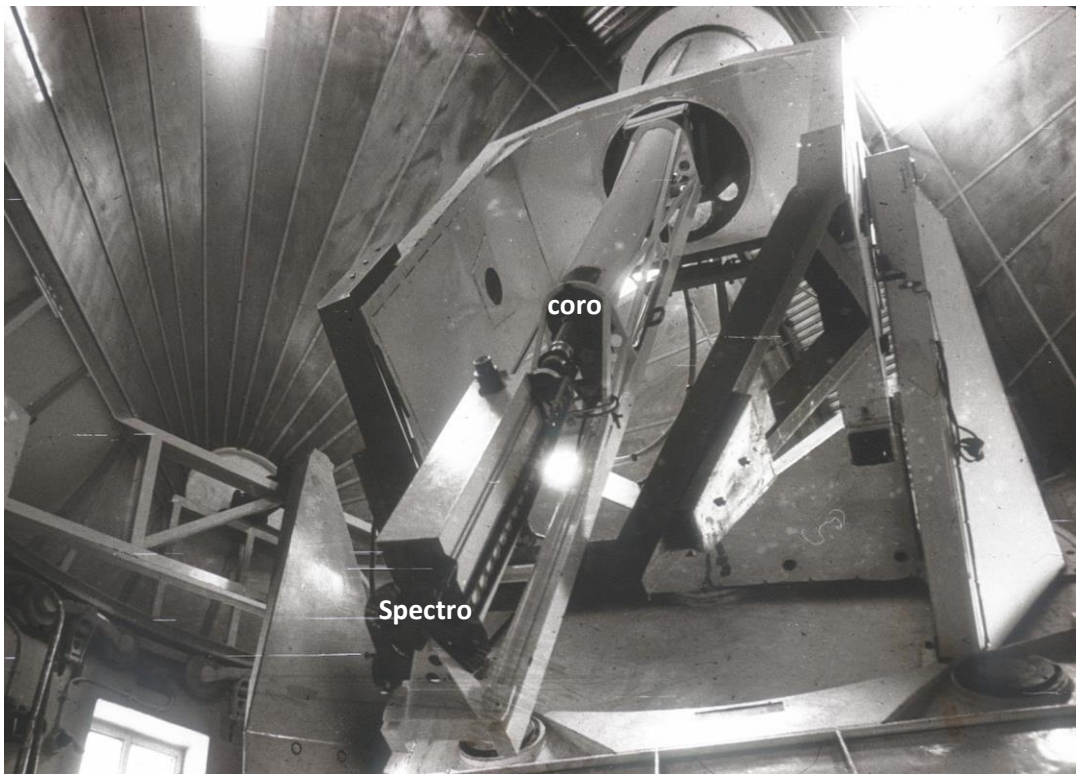

Figure 31: spectro-coronagraph of 15 cm and 1.5 m focal length, and the spectrograph of 52 cm focal length, put on the mount of the turret dome after removal of the refractor (credit J.P. Rozelot).

**CONCLUSION**

The 1950 decade (from 1947 to 1961) was marked by a decisive development of the Pic du Midi infrastructure, under the direction of Rösch : the high-voltage power line, the funicular, the cable car. Without this equipment, the Wilson chambers of the teams of physicists, using cosmic rays as a tool for nuclear physics, would not have been able to operate at the Pic. Year 1956 marked the beginning of spectroscopy of the solar surface (photosphere, chromosphere, flares, sumspots), under the auspices of Michard, which Mouradian later continued on the turret dome refractor. In parallel, several spectrographs designed for the corona (above the limb) succeeded the initial Lyot instruments. The experience acquired by researchers on the large spectrographs of Meudon (spectroheliograph, magnetograph, solar tower) and Pic du Midi favoured the design of the THEMIS telescope in the Canary Islands, which has been operating for 25 years. This know-how may be extended in the future to the large European telescope, a running project planned for the 2030s.

**REFERENCES**

Demarcq, J., Hugon, M., Rösch, J., Treillis, M., 1965, « un coronographe achromatique amélioré pour la spectrographie de la couronne solaire », *CRAS*, 261, 4629

Hugon, M., Rösch, J., Treillis, M., 1963, « l'observation depuis le sol de la couronne d'émission », *IAUS*, 16, 215

Lyot, B., 1931, « Etude de la couronne solaire en dehors des éclipses », *ZA*, 5, 73

Lyot, B., 1932, « la photographie de la couronne solaire hors des éclipses et son étude au spectrohéliographe », *l'Astronomie*, 272

Malherbe, J.-M., Roudier, Th., Moity, J., Mein, P., Arnaud, J., Muller, R., 2007, "Spectropolarimetry with liquid crystals", *Mem. S. A. It.*, 78, 203.

Mein, P., 1980, "Multichannel Subtractive Double Pass Spectrograph", *Proceedings of the Japan France Seminar on Solar Physics,* Moryama and Hénoux editors*,* 285.

Mein, P., Mein, N., 2020, "Raymond Michard and his solar physics group at Paris Meudon observatory", *JAHH*, 23, 582

Michard, R., Mouradian, Z., Semel, M., 1961, « champs magnétiques dans un centre d'activité solaire avant et pendant une éruption », *An. Ap.*, 24, 54

Mouradian, Z., Chauveau, F., Colson, F., Darré, G., Kerlirzlin, P., Olivieri, G., 1980, « The new solar spectrograph at Pic du Midi Observatory", *Proceedings of the Japan France Seminar on Solar Physics,* Moryama and Hénoux editors*,* 271.

Noens, J.C., Pageault, J., Ratier, G., 1984, « Measuring electron density in coronal active regions ; a multichannel coronagraph with photoelectric spectrograph", *Solar Phys.*, 94, 117

Rösch, J., 1963, « l'observatoire du Pic du Midi », *revue de l'enseignement supérieur*, 103.

Roudier, T., Malherbe, J.M., Rozelot, J.P., Mein, P., Muller, R., 2021, « Five decades of solar research at the Pic du Midi turret dome", *JAHH*, 24, 585

Rozelot, J.P., 1972, « Analyse des renforcements coronaux à travers quelques acquisitions spectroscopiques récentes des émissions monochromatiques du fer ionisé (X à XV) », *Solar Phys*., 22, 88-113.

Rozelot, J.P., Despiau, R., 1972, « Adaptation de la caméra électronique au coronographe », *CRAS*, 275, 613